%% file: acl_latex.tex
\documentclass[11pt]{article}

\usepackage[preprint]{acl}

\usepackage{times}
\usepackage{latexsym}
\usepackage{musixtex}
\usepackage[T1]{fontenc}

\usepackage[utf8]{inputenc}

\usepackage{microtype}

\usepackage{inconsolata}
\usepackage{amsmath} 
\usepackage{hyperref}
\usepackage{url}
\usepackage{multirow}
\usepackage{microtype}
\usepackage{url}
\usepackage{lilyglyphs}
\usepackage{booktabs}
\usepackage{wrapfig}
\usepackage{graphicx}
\usepackage{caption}
\usepackage{url}
\usepackage{subcaption}
\usepackage{comment}
\usepackage{longtable}
\usepackage{makecell}
\usepackage{siunitx}
\usepackage[table]{xcolor}
\usepackage{booktabs}
\usepackage{tabularx}
\usepackage{longtable}
\usepackage{multirow}

\usepackage{makecell}
\usepackage{siunitx}  
\usepackage{adjustbox}  

\usepackage{mycommands}

\definecolor{darkblue}{rgb}{0, 0, 0.5}
\hypersetup{colorlinks=true, citecolor=darkblue, linkcolor=darkblue, urlcolor=darkblue}

\makeatletter
\renewcommand{\sectionautorefname}{\S\@gobble}
\renewcommand{\sectionautorefname}{\S\@gobble}
\renewcommand{\subsectionautorefname}{\S\@gobble}
\renewcommand{\sectionautorefname}{\S\@gobble}
\renewcommand{\subsectionautorefname}{\S\@gobble}
 \renewcommand{\appendixautorefname}{\S\@gobble}
\makeatother

\definecolor{purp}{HTML}{791f87}
\definecolor{highlight}{RGB}{255, 255, 0}
\definecolor{bottlegreen}{rgb}{0.0,0.42,0.31}
\definecolor{bblue}{rgb}{0.0, 0.6, 0.8}

\newcommand{\benchmark}{\textsc{BASS}}

\DeclareFontFamily{U}{musix}{}
\DeclareFontShape{U}{musix}{m}{n}{<-> s*[1.00] musix11}{}
\newcommand{\mathbassclef}{%
  \text{%
    \vphantom{A}%
    \raisebox{.55\height}[0pt][0pt]{\usefont{U}{musix}{m}{n}\symbol{73}}%
  }%
}

\title{BASS $\text{\mathbassclef}$: Benchmarking Audio LMs for \\ Musical Structure and Semantic Reasoning} 

\author{
\textbf{Min Jang}$^{1,\spadesuit}$,
\textbf{Orevaoghene Ahia}$^{1,\spadesuit}$,
\textbf{Nazif Tamer}$^1$ \\
\textbf{Sachin Kumar}$^2$,
\textbf{Yulia Tsvetkov}$^1$,
\textbf{Noah A. Smith}$^{1,3}$ \\
$^1$University of Washington,
$^2$The Ohio State University,
$^3$Allen Institute for AI \\
$^\spadesuit$Denotes equal contribution
}

\begin{document}

\maketitle
\input{abstract}

\input{01_introduction}

\input{02_melody}

\input{03_experimental_setup}
\input{04_Results}

\input{06_related_work}
\input{07_conclusion}


\bibliography{custom}

\newpage
\appendix

\section*{Appendix}
\label{sec:appendix}
\input{appendix}


\end{document}

%% file: abstract.tex
\begin{abstract}
Music understanding is a complex task that often requires reasoning over both structural and semantic elements of audio. We introduce \benchmark{}, designed to evaluate music understanding and reasoning in audio language models across four broad categories: structural segmentation, lyric transcription, musicological analysis, and artist collaboration.  \benchmark{} comprises 2658 questions spanning 12 tasks, 1993 unique songs and covering over 138 hours of music from a wide range of genres and tracks, crafted to assess musicological knowledge and reasoning in real-world scenarios. We evaluate 14 open-source and frontier multimodal LMs, finding that even state-of-the-art models struggle on higher-level reasoning tasks such as structural segmentation and artist collaboration, while performing best on lyric transcription. Our analysis reveals that current models leverage linguistic priors effectively but remain limited in reasoning over musical structure, vocal, and 
musicological attributes. \benchmark{} provides an evaluation framework with widespread applications in music recommendation and search and has the potential to guide the development of audio LMs.
\footnote{
Correspondence: \scriptsize \texttt{minjang, oahia@cs.washington.edu}\\
\textbf{Website}: \scriptsize \url{https://minjang10.github.io/bass-website/} \\
\textbf{Code}: 
\url{https://github.com/minjang10/bass_music_benchmark}.\\
\textbf{Data}: \scriptsize \url{https://huggingface.co/datasets/oreva/bass_music_benchmark}
}

\end{abstract}

%% file: 01_introduction.tex
\section{Introduction}
There has been significant progress in developing audio language models capable of understanding and reasoning across diverse audio domains, including speech, environmental sounds, and music \citep{goel2025audio, ghosh2025audio, gemini2023multimodal, kimiteam2025kimiAudio, xu2025qwen3Omni, xie2025audioreasoner, wijngaard2025audsemthinker}. Music is particularly challenging due to its rich layered structure, encompassing harmony, rhythm, form, features often absent in simpler audio domains. Accurate music understanding capabilities could benefit listeners, scholars, performers, producers, composers, and media companies by enabling open-ended, natural-language interactions with the content in individual music samples or large-scale collections.

\begin{figure}[t!]
    \centering
    \includegraphics[width=\columnwidth]{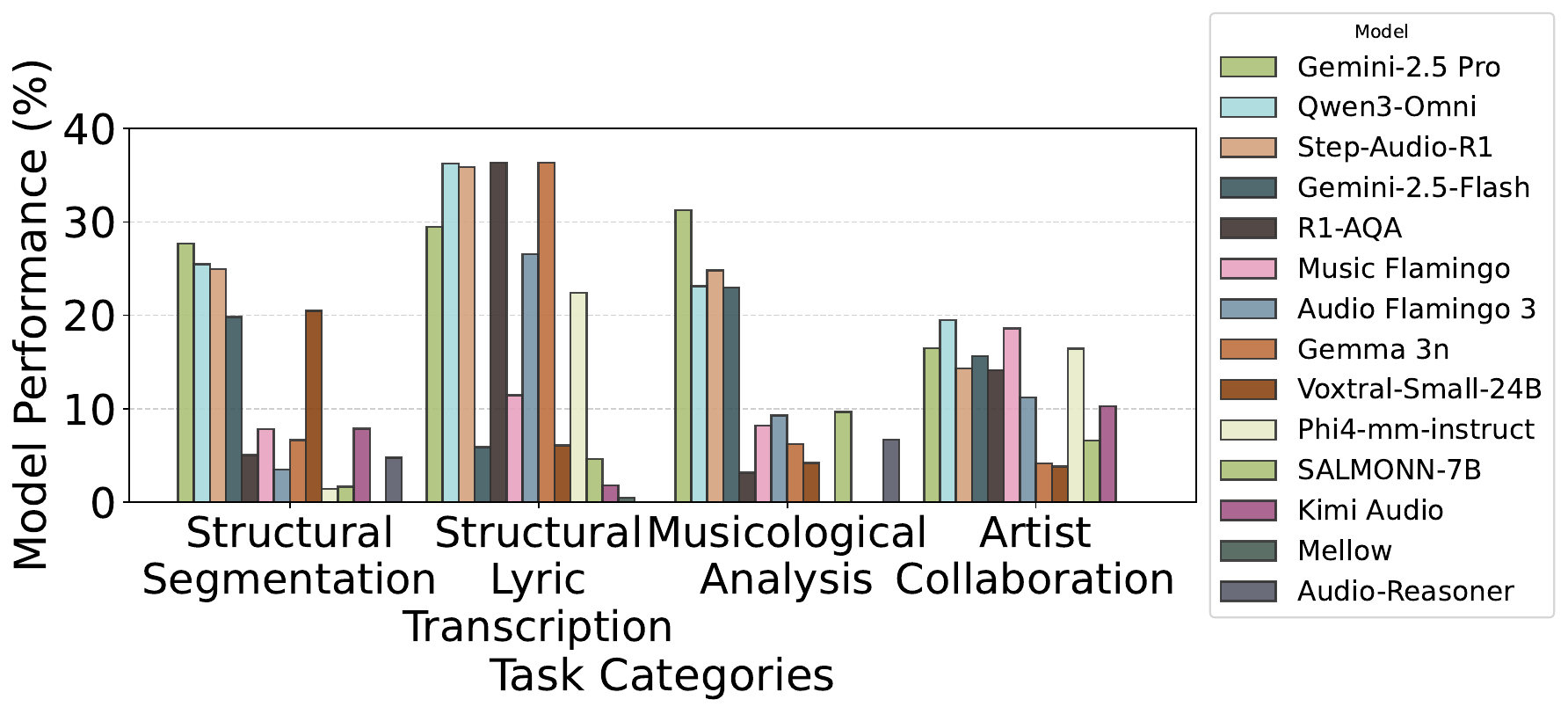}
    \caption{Average performance of audio LMs across all task categories in \benchmark{} (
    the upper bound is 100\% in each case). Top-performing models achieve their highest scores on lyric transcription and their lowest on artist collaboration tasks.
    } 
    \label{fig:normalized_categories_spider_plot}
\end{figure}

\begin{figure*}[hbtp]
  \centering
  \resizebox{0.9\textwidth}{!}{
  \includegraphics[width=\linewidth]{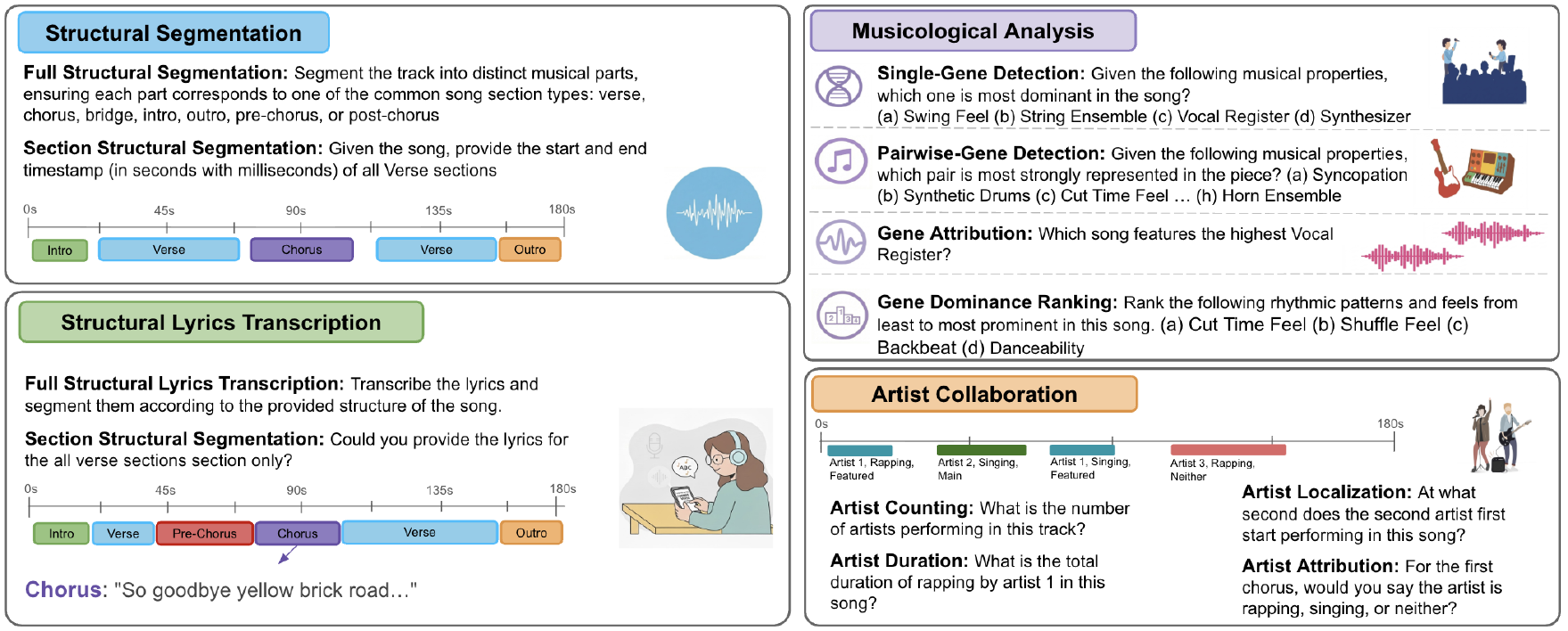}}
    \caption{Overview of the tasks included in \benchmark{} Benchmark, designed to evaluate audio LMs on musical understanding requiring long-term structure or hierarchical reasoning. It contains twelve tasks across four categories: structural segmentation, structural lyrics transcription, musicological analysis, and artist collaboration. } 
  \label{fig:benchmark}
\end{figure*}

Existing benchmarks mostly focus on music captioning or lyrics captioning \citep{manco2023thesong, agostinelli2023musiclmgeneratingmusictext} or cover broad audio domains, including  but not specializing in music \citep{ahia2025blabbrutallylongaudio, sakshi2025mmau, kumar2025mmau, ma2025mmar, huang2025dynamic_superb_phase2} and hence fail to evaluate deep musical reasoning across diverse contexts. MuChoMusic \citep{weck2024muchomusic} addresses some of these gaps but relies entirely on captioning datasets \citep{manco2023thesong, agostinelli2023musiclmgeneratingmusictext}, uses mostly short songs ($\sim$2 minutes), and lacks tasks requiring long-term structure or hierarchical reasoning.

We present \benchmark{},
a benchmark that solely evaluates music understanding and reasoning in audio LMs across 12 tasks under four broad categories: structural segmentation, lyric transcription, musicological analysis, and artist collaboration. Our task selection process is guided by practical use cases of LMs to answer real-world music-related questions most naturally posed in  language. Structural segmentation includes tasks that require identifying and localizing musical sections within a recording. Lyric transcription evaluates a model’s ability to transcribe song lyrics at varying granularity levels. Musicological analysis focuses on identifying and reasoning about abstract musicological attributes. 
Artist collaboration includes tasks that require identifying, attributing, and temporally localizing multiple vocalists within a song, often involving fine grained temporal reasoning.

Using  \benchmark{}, we evaluated 14 open-source and frontier multimodal LMs \autoref{sec:experimental_steup}. Our results indicate that even frontier models such as Gemini 2.5 Pro and strong open-source models like Qwen3-Omni struggle substantially on several tasks. As shown in \autoref{fig:normalized_categories_spider_plot}, top performing models achieve their best results on lyric transcription and perform worst on artist collaboration tasks. We conduct an in-depth analysis of model performance across all tasks and our results suggest that current audio LMs are much stronger on tasks where they can leverage linguistic priors, but remain limited in higher level reasoning over musical structure, vocal style, and multi-artist profiles.
\benchmark{} provides a
challenging testbed for evaluating music understanding and reasoning in audio LMs, highlighting their current strengths and their limitations. We will publicly release the dataset alongside all evaluation code associated with the benchmark upon acceptance.

%% file: 02_melody.tex
\section{\benchmark{} Benchmark} 

\benchmark{} is designed to evaluate music understanding and reasoning in AI systems. It consists of 2,658 question-answer pairs across 12 diverse tasks under four broad categories: a) structural segmentation
b) lyric transcription, c) musicological analysis, and d)
artist collaboration. Unlike prior music benchmarks such as \citet{weck2024muchomusic}, which primarily evaluate models using multiple-choice questions with four distractor options, \benchmark{} incorporates both open-ended and multiple-choice questions, with distractor options increased to eight choices in some cases, reducing the likelihood of models solving tasks by guessing and enables evaluation of deeper musical reasoning.
All music recordings across every task were sourced from YouTube.
\footnote{Due to licensing restrictions, we will provide YouTube links to the tracks rather than distributing the audio directly, allowing legal access while respecting copyright.} Each sample was manually verified to confirm song accuracy and ensure high audio fidelity.
The average duration of audio clips in the dataset is 4.16 minutes, spanning a range of 49.06 seconds to 19.44 minutes.
 \autoref{tab:data_stats} in the Appendix 
 provides detailed statistics for \benchmark{}, and below we describe the rigorous annotation procedure for each task in depth. 

\subsection{\BroadCategory{Structural Segmentation}}
Structural segmentation, well researched in the field of music information retrieval, requires musical knowledge and understanding of key structural sections in musical audio \cite{paulus-2010-state,ehmann2011music}. Given a music recording, the objective is to identify important structural elements and temporally segment the recording according to these elements. This capability is  essential in consumer applications; streaming platforms like Spotify use structural information to improve recommendations, playlisting, and highlight extraction \citep{Montecchio_2020, MusicTomorrow2025, choi2016playlistgenerationalgorithmsusing}. Within this category, we include two tasks: \textbf{full} and \textbf{section} structural segmentation. We source data for this task from Harmonix Set \citep{nieto-2019-harmonix}, a dataset with structural annotations for pop songs. Originally, it  contained 61 structural section labels, with 5 very common sections  (Intro, Verse, Chorus, Bridge, or Outro) and a long tail of 56 sections. 

\noindent \Subcategory{Full Structural Segmentation} (FSS) requires the models to segment the entire song into its major sections and return start–end timestamps for each. 

\noindent \Subcategory{Section Structural Segmentation} (SSS) targets a single section type per song (e.g., chorus), and either returns timestamps for all occurrences of that section type (note that sections often repeat) or returns a specific section (e.g., second Chorus) in the recording. 


\subsection{\BroadCategory{Structural Lyric Transcription}}
Structural lyric transcription extends standard lyric transcription  \citep{wang2022catch, mesaros2008automatic} 
 by requiring models to transcribe lyrics within each structural section of a song.
 We use Genius \footnote{\url{https://genius.com/}} Lyrics due to its collaborative annotation process, where transcriptions and segmentations of popular tracks are refined by hundreds of contributors, ensuring higher reliability. This task is more challenging than structural segmentation alone, as models must both segment the song and correctly align generated lyrics to each section. Successful performance enables applications such as content indexing, music recommendation, and more reliable automated, and localized, lyric transcription. Within the structural lyric transcription category, we again have \textbf{Full} and \textbf{Section} variants. 

\noindent \Subcategory{Full Structural Lyrics Transcription} (FSLT): the model is tasked with transcribing the lyrics of each section in a song, using as the gold-standard reference 
lyrics that were crawled from Genius Lyrics that contain common sections (Intro, Verse, Chorus, Bridge, Outro). We also include additional sections, such as Pre-Chorus and Post-Chorus, which frequently appear. To reduce ambiguity, each question is templated: all relevant sections are listed in the prompt, and the model is asked to generate the corresponding lyrics for each section.

\noindent \Subcategory{Section Structural Lyrics Transcription}(SSLT): the model is tasked with transcribing
the lyrics of a specific section (e.g., Bridge). The task includes two types of questions: all section-level SSLT and single section-level SSLT. For all section-level SSLT, the model is given a section name and must transcribe all instances of that section in the song. For single section-level SSLT, a specific instance is provided (e.g., the second verse), and the model must transcribe only that instance.



\subsection{\BroadCategory{Musicological Analysis}} \label{sec:musicological}
Musicological analysis examines the structural and expressive elements of music—rhythm, harmony, melody, timbre, vocals, and lyrics—which provide explainable foundations for other common tasks like music classification and recommendation \citep{Castelluccio2006, oramas2025mgphot}. In this benchmark, we evaluate how well audio language models can identify musicological attributes referred to as \textit{genes}: specific, measurable attributes of songs such as the prevalence of drum use, electric guitar distortion, sad lyrics, or emphasis on rhythmic groove.

We derive these genes from the MGPHot dataset \citep{oramas2025mgphot}, which comprises 58 musical attributes organized into seven categories: rhythm, instrumentation, sonority, harmony, compositional focus, vocals, and lyrics. These attributes were annotated by professional musicologists as part of the Music Genome Project \citep{Castelluccio2006} and cover 21,320 songs that appeared on the Billboard Hot 100 at least once between 1958 and 2022. Musicologists rated the dominance of each gene on a 0–5 scale which was then translated to a 0.0-1.0 scale, where 1.0 indicates high dominance and 0.0 indicates absence. Using these professional annotations, we curated four challenging tasks for audio LMs: (a) \textbf{Single-Gene Detection}, (b) \textbf{Pairwise-Gene Detection}, (c) \textbf{Gene Attribution}, and (d) \textbf{Gene Dominance Ranking}.

\noindent \Subcategory{Single-Gene Detection} (SGD):  the model is tasked to identify the most dominant musicological attribute in a given recording from four options, given each gene’s name and description. Questions are constructed from songs where the gene is strongly expressed ($\geq 0.6$). Distractors are chosen from attributes with values positively correlated with the ground truth but negatively correlated with each other to avoid overly similar options within the same gene category (e.g., "Synthesizer" and "Synth Timbre").

\noindent \Subcategory{Pairwise-Gene Detection} (PGD): requires an audio LM to select the pair of musicological attributes most prominently expressed in a given recording, given eight musicological attributes and their descriptions. This task is more challenging than SGD, as ground-truth pairs are drawn from attributes that are typically negatively correlated e.g (Acoustic Guitar and Synthetic Sonority), yet questions are based on songs where both attributes are strongly expressed. Distractors are selected from genes positively correlated with at least one ground-truth attribute but negatively correlated with all other distractors.

\noindent \Subcategory{Gene Attribution} (GA)
moves beyond single-song analysis to evaluating musicological attributes across four recordings. The model is given a musicological attribute and its description and must select which of four recordings expresses this attribute most prominently. 
Eligible questions require the ground-truth song to have a musicological attribute value greater than 0.8, while distractor songs must have values below 0.5 to ensure sufficient contrast. The models are provided with a single audio input, which contains each song concatenated with 5 seconds of silence between each song. The model is expected to output the index of the recording in which the attribute is most prominently expressed (1 for the first recording, 2 for the second, etc.)
Evaluation is done by measuring accuracy over the four-way choice task.

\noindent In \Subcategory{Gene Dominance Ranking} (GDR), the model is given four musicological attributes for each song and must rank them in ascending order based on their prominence. To avoid ambiguity, the attribute values for each song are designed to be uniformly spaced. The models must produce the rankings in the exact correct order to be considered correct. No credit is given for partially correct outputs. 
\subsection{\BroadCategory{Artist Collaboration}}
Several studies have analyzed the musical characteristics of collaborative songs \citep{5d6ef325-618e-4676-bd8d-9e2c7de417fa,KomanieckiRobert2017Acfi,Duguay2022}, but none have analysed how effectively audio LMs understand these properties. In popular music, songs often feature multiple vocalists, such as a main and a featured artist, or are performed by bands with several members. We design this task to evaluate how well audio LMs can understand and reason about artist roles (main or featured), vocal delivery types (rap or singing), and their temporal distribution across different formal sections of a song (e.g., introduction, verse, chorus). This includes the model’s ability to localize when each artist or vocal delivery type occurs throughout the track.

Questions and ground truth were curated from fine‑grained timed annotations in the Collaborative Song Dataset (CoSoD; \citealp{DuguayManceyDevaney2023}), which includes 331 multi-artist songs annotated with formal structure and vocal production features like layering, panning, and delivery type. We designed these tasks to evaluate core dimensions of artist behavior and vocal performance-- specifically, artist presence, timing, vocal style, and section-specific artist roles. They capture the most relevant dimensions for analyzing artist contributions in collaborative music, which can support real-world applications such as music recommendation, and  automatic metadata generation. Within this category, there are four tasks, discussed in turn.

\noindent \Subcategory{Artist Counting} consists of five subtasks. \textit{Standard artist counting} asks the model to report the total number of artists credited in a song, regardless of their role or level of contribution, including both primary and featured artists. 
\textit{Featured artist counting} focuses on songs with more than two featured artists and asks the model to report the total number of featured artists on the track. Featured artists are additional performers who contribute to the song but are not the primary lead artist. 
\textit{Vocal delivery counting} requires the model to count the number of artists performing with a specific vocal delivery style,(rap, singing) in a single recording. \textit{Section artist} counting involves counting artists within a specific section of a song (e.g., introduction, verse, chorus) or across multiple sections, including negative cases such as empty or nonexistent sections. \textit{Temporal artist counting} asks for the number of artists within a given time range or across multiple timestamps, including out-of-bounds intervals where the count is zero. 
For each of the artist counting subtasks, the model is expected to output the exact number of artists, and performance is evaluated using exact-match accuracy between the model’s output and the ground-truth count.

\noindent \Subcategory{Artist Duration} includes four subtasks. 
\textit{Target artist duration} requires the model to report the total duration a specific artist performs in the song, measured in seconds. The artist is identified by their order of appearance rather than by name, to ensure fairness for models that may not have prior exposure to the song. \textit{Vocal delivery duration} asks the model to report the total time, in seconds, during which a specific vocal delivery type (e.g., rap, singing) occurs in the song, regardless of which artist performs it. Artist delivery duration combines the previous two subtasks, requiring the model to report the total time, in seconds, that a specific artist performs using a specific vocal delivery style (e.g., rap, singing). Finally, \textit{Section duration} tasks the model with estimating the total duration, in seconds, of a particular section type (e.g., all verse sections). Evaluation is done using exact-match accuracy, where the model's output must match the ground-truth duration within a tolerance of $\pm 3$ seconds.

\noindent \Subcategory{Artist Localization} requires the model to provide the start timestamp, in seconds, of when a specific artist first appears in the song. The specific artist is identified by their order of appearance in the song. A 3 second time window from the ground truth timestamp is used to evaluate each model's output. Evaluation is done using exact-match accuracy, where the model’s output is considered correct if it falls within a ±3 second window of the ground-truth timestamp.

\noindent \Subcategory{Artist Attribution} includes four subtasks. \textit{Vocal style comparison} asks the model to determine whether the vocal styles of different artists within the same song are the same or different. For example, the task might ask whether the first and second artists in the song are both rapping. 
 \textit{Sectional vocal style} requires identifying whether an artist in a specific section is singing, rapping, or neither. \textit{Sectional artist role} tasks the model with determining an artist’s role (main, featured, or neither) within a section. For example, the model might be asked whether a particular artist is the main performer or a featured artist in the chorus. Finally, \textit{Temporal vocal style} asks the model to identify the vocal style of an artist within a given start and end timestamp.  Evaluation for all artist attribution tasks is done using exact-match accuracy.

%% file: 03_experimental_setup.tex
\section{Experimental Setup}\label{sec:experimental_steup}
\input{tables/full_music_normalized_scores}
\label{experimental-setup}
\paragraph{Models}

We benchmark a wide range of audio LMs.
Among open-source models, we include Qwen3-Omni \citep{xu2025qwen3Omni}, Audio Flamingo 3 \citep{goel2025audio}, Music Flamingo \citep{ghosh2025musicflamingoscalingmusic}, Step-Audio-R1 \citep{tian2025stepaudior1technicalreport}, R1-AQA \citep{li2025reinforcementlearningoutperformssupervised}, Audio Reasoner \citep{xie2025audioreasoner}, Phi 4 Multimodal Instruct \citep{microsoft2025phi4minitechnicalreportcompact}, Voxtral \citep{liu2025voxtral}, SALMONN-7B \citep{tang2024salmonngenerichearingabilities}, Mellow \citep{deshmukh2025mellowsmallaudiolanguage}, Kimi-Audio \citep{kimiteam2025kimiAudio}, and Gemma 3 \cite{gemmateam2025gemma3technicalreport}.  The closed-source models include Gemini 2.5 Pro, Gemini 2.5 Flash (with thinking), and Gemini 2.5 Flash (without thinking) \citep{gemini2023multimodal}.\footnote{We do not evaluate OpenAI models due to cost limitations; in fact, processing audio with these models is roughly 16 times more expensive than using Gemini 2.5 Pro. }

\paragraph{Evaluation Strategy}
In all our experiments, audio LMs receive a text prompt (instruction) and an audio file as input and produce text as output. 
For each task, the authors handcraft 10  natural language prompts that are paraphrases of each other, meaning they all ask the same underlying question in slightly different ways. 
Across the audio samples for each task, we maintain a balanced mixture of these paraphrases to promote diversity. To ensure models truly process the audio and avoid relying on shortcuts, 
we evaluate all tasks using free-form generation, except for Musicological Analysis questions, in which we use a multiple-choice format as explained in \S\ref{sec:musicological}. The performance on multiple-choice questions is determined using majority voting over four runs.

\paragraph{Metrics}\label{sec:metrics_description}
For Structural Lyric Transcription, we use word error rate (WER). For questions involving multiple sections, reference and predicted sections are optimally matched, and the final WER is the average across all sections. For Structural Segmentation, performance is measured using intersection over union (IoU), where an overlap counts as an intersection only if the predicted section labels match the reference.
For all other tasks, we use exact match accuracy (EMA). For questions requiring timestamps or durations, predictions within a 3-second offset are considered correct.

Aggregating the metrics across tasks poses two challenges. First, metrics differ in their direction of optimality: WER is minimized, while EMA is maximized. Furthermore, WER is unbounded.\footnote{Although WER for standard ASR tasks tend to be between 0.0 and 1.0, because musical recordings are long-form audio and contain much more background noise, we found that many models achieved a WER much higher than 1.0, particularly due to excessive insertions, making WER unbounded.} Second, while EMA lies between 0.0 and 1.0, naively averaging EMAs across different tasks is invalid due to the tasks having different random chance accuracy. Hence, we invert WER so that higher scores corresponds to better performance, $\text{IWER} = \frac{1}{1 + \text{WER}}$, 
bounding the score between 0.0 and 1.0. To address the second issue, we normalize EMA to represent models' improvement over random guessing. Specifically, we report $\text{EMA}_{N} = \frac{\text{EMA} - R}{1 - R}$
where $R$ is random guessing accuracy.  This gives an upper bound of 1.0 (but $\text{EMA}_N$ can be negative if a system performs worse than random guessing on average). All our results are reported in percentages (\%). 

%% file: tables/full_music_normalized_scores.tex
\begin{table*}[t]
\rmfamily
\centering
\small
\setlength{\tabcolsep}{3pt}
\adjustbox{max width=\textwidth}{
\begin{tabular}{l l
S[table-format=2.2] S[table-format=2.2] S[table-format=2.2] S[table-format=2.2]
S[table-format=2.2] S[table-format=2.2]
S[table-format=2.2] S[table-format=2.2] S[table-format=2.2] S[table-format=2.2]
S[table-format=2.2] S[table-format=2.2] S[table-format=2.2]
}

\toprule
 & & \multicolumn{4}{c}{\textbf{\BroadCategory{Musicological Analysis}}} 
 & \multicolumn{2}{c}{\textbf{\BroadCategory{Structural Segmentation}}}
 & \multicolumn{4}{c}{\textbf{\BroadCategory{Artist Collaboration}}}
 & \multicolumn{2}{c}{\textbf{\BroadCategory{Lyrics Transcription}}}  & {\textbf{Average}} \\
\cmidrule(lr){3-6} \cmidrule(lr){7-8} \cmidrule(lr){9-12} \cmidrule(lr){13-14}

\textbf{Model} & \textbf{Size}
& {SGD} & {PGD} & {GDR} & {GA} 
& {Full} & {Section} 
& {Counting} & {Duration} & {Localization} & {Attribution} 
& {Full} & {Section} & \\

\midrule
Gemini-2.5 Pro & - & \textbf{64.98} & 10.40 & 17.80 & 31.90 & \textbf{32.34} & \textbf{22.98} & 25.77 & 8.25 & 3.12 & 28.84 & \textbf{42.63} & 29.45 & \textbf{26.54} \\
Step-Audio-R1 & 33B & 43.88 & 8.89 & 14.62 & \textbf{31.91} & 32.19 & 17.73 & 19.34 & 6.42 & \textbf{7.81} & 23.74 & 21.86 & 35.89 & 22.24 \\
Qwen3-Omni & 30B & 47.86 & 1.27 & 11.51 & 31.87 & 28.88 & 22.07 & 18.54 & \textbf{9.78} & 9.38 & \textbf{40.32} & 25.53 & \textbf{36.25} & 23.61 \\
Gemini-2.5-Flash & - & 43.48 & 4.06 & \textbf{20.90} & 23.50 & 24.47 & 15.15 & 20.40 & 7.66 & 3.12 & 31.44 & 5.13 & 5.89 & 17.10 \\
Music Flamingo & 8B & 32.34 & 4.07 & -4.34 & 0.67 & 13.07 & 2.59 & 33.54 & 4.27 & 4.69 & 31.88 & 15.72 & 11.47 & 12.50 \\
R1-AQA & 8.2B & 12.04 & -2.98 & -4.35 & 7.87 & 5.91 & 4.16 & \textbf{35.03} & 0.92 & 0.00 & 20.51 & 40.05 & 36.35 & 12.96 \\
Audio Flamingo 3 & 8B & 25.60 & 8.01 & -4.34 & 7.87 & 7.03 & 0.01 & 21.92 & 1.84 & 0.00 & 21.13 & 17.59 & 26.57 & 11.10 \\
Gemma 3 & 5B & 13.63 & 3.30 & -1.19 & 9.22 & 8.92 & 4.35 & 12.25 & 0.30 & 1.56 & 2.47 & 40.05 & 36.35 & 10.93 \\
Voxtral-Small & 24B & 16.82 & 5.42 & 5.15 & -10.67 & 32.16 & 8.84 & 9.09 & 2.76 & 0.00 & 3.42 & 26.45 & 6.08 & 8.79 \\
SALMONN-7B & 7B & 0.50 & \textbf{37.26} & -4.03 & 4.93 & 3.18 & 0.12 & 27.32 & 0.61 & 0.00 & -1.48 & 11.61 & 4.63 & 7.05 \\
Phi4-mm-instruct & 5.5B & -4.96 & 1.27 & 5.15 & -4.93 & 2.71 & 0.16 & 23.59 & 3.66 & 0.00 & 38.51 & 4.99 & 22.45 & 7.71 \\
Audio-Reasoner & 8.4B & 11.64 & 5.37 & 1.98 & 7.80 & 6.63 & 2.90 & 18.47 & 3.06 & 0.00 & -59.03 & 2.46 & 0.00 & 0.00 \\
Kimi Audio & 7B & 16.82 & 0.55 & -4.35 & -33.33 & 13.06 & 2.64 & 19.12 & 0.92 & 0.00 & 20.96 & 5.17 & 1.80 & 3.61 \\
Mellow & 167M & -32.94 & -3.70 & -4.35 & -33.33 & 0.00 & 0.00 & 0.00 & 2.14 & 0.00 & -62.50 & 53.50 & 0.55 & -6.72 \\
\bottomrule
\end{tabular}
}
\caption{Normalized performance of evaluated models on \benchmark{} across musicological analysis (Single-Gene Detection—SGD, Pairwise-Gene Detection—PGD, Gene Dominance Ranking—GDR, and Gene Attribution-GA) , structural segmentation, artist collaboration, and lyrics transcription tasks. Each score upper bound is 100\% and bold values highlight the highest value per task. The normalization procedure is described in \autoref{sec:metrics_description}. The raw scores are provided in \autoref{tab:raw_scores_full_music} in the Appendix. }
\label{tab:normalized_scores}
\end{table*}

%% file: 04_Results.tex
\section{Results and Discussion}\label{sec:results}
\paragraph{Benchmarking Results}
\begin{figure*}[t] 
    \centering
    \includegraphics[width=0.95\textwidth]{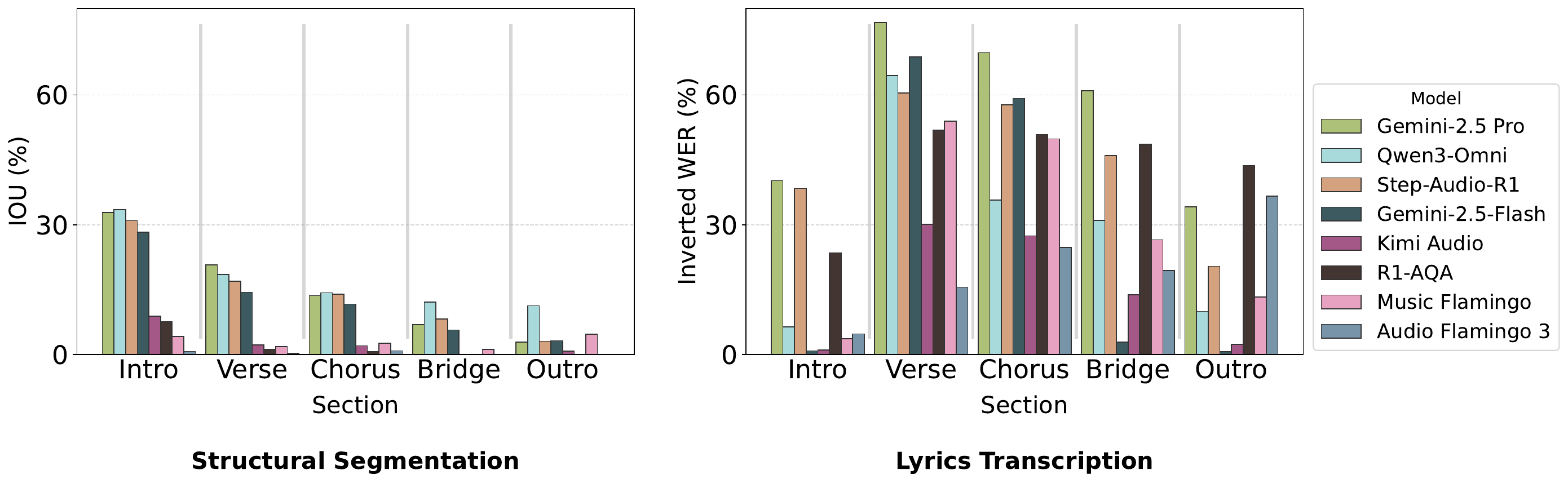}
    \caption{Section-wise performance for Structural Segmentation and Lyric Transcription (higher is better). Segmentation performance declines over time, while transcription performance (IWER) is lowest for the \textit{Intro} compared to \textit{Verse} and \textit{Chorus} sections.}
    \label{fig:structural_tasks}
\end{figure*}

In \autoref{tab:normalized_scores}, we present results containing normalized scores across all tasks in \benchmark{} for all models.  
The raw scores are provided in \autoref{tab:raw_scores_full_music} in the Appendix.
On average, performance across all models remains below 30\%; the best is  Gemini-2.5 Pro at 26.54\% and Qwen3-Omni following closely at 23.61\% and Step-Audio-R1 at 22.24\% . Notably, all of these models are trained with hybrid thinking modes, and these evaluations were conducted with thinking enabled. 
Overall, models perform best at lyric transcription, while artist collaboration tasks prove to be the most challenging. The performance gap suggests that current audio LMs are stronger on tasks where they can leverage linguistic priors, but are limited in higher level quantitative and temporal reasoning over musical structure, vocal style, and multi-artist profiles. 

Below, we present a detailed discussion of our results, framed around the research questions that guided our study. Specifically, we examine how audio LMs understand and reason about musical structure and lyrics, how they handle artist roles and vocal styles, and how their performance varies across different genres. 

\paragraph{How well do audio LMs understand and reason about musical structure?}
On structural tasks like structural segmentation and lyrics transcription, we observe that models often perform better 
when the task requires 
temporal localization across full songs rather than isolated sections. We suspect that reasoning sequentially over earlier sections allows models to use contextual cues from preceding segments to better identify later sections. However, even when models perform temporal localization across full songs (see \autoref{fig:structural_seg_heatmap}), they frequently get confused between acoustically similar sections, such as verses and choruses. We speculate that this confusion is due to audio LMs not being explicitly trained to
understand musical structure across long contexts. This is evident in our analysis of model performance across different song sections (see \autoref{fig:structural_tasks}), where models are generally more capable of localizing content in sections that occur earlier in the song, such as intros and first verses than later sections such as bridges and outros.

\paragraph{Does providing an ASR transcript improve audio LMs’ structural lyric transcription?} 
General‑purpose automatic speech recognition models \citep{10.5555/3618408.3619590} show strong generalization to automatic lyric transcription \citep{wang2023adapting, zhuo2023lyricwhiz, ou2022transfer, cifka-2024-jam-alt, 10.5555/3618408.3619590}.
However, they also note some errors in the transcriptions. This motivates us to explore cascaded approaches to structural lyric transcription.  Combining linguistic cues from the transcript with acoustic information from the audio could help the model determine section boundaries and align lyrics more accurately.
We first use Whisper to transcribe the song and then feed the resulting transcript, alongside the audio, back into the audio LM as a guide to perform the structural transcription task 
As shown in \autoref{tab:cascaded_vs_audio}, we find that using a cascaded Whisper transcript actually \emph{reduces} audio LMs’ lyrics transcription performance compared to audio-only transcription. We speculate that  transcripts can mislead the model and introduce ASR errors, hurting performance on sections that require precise temporal localization.


\input{tables/cascaded_lyrics_transcription}

\begin{figure}[h]
    \centering
    \includegraphics[width=0.8\columnwidth]{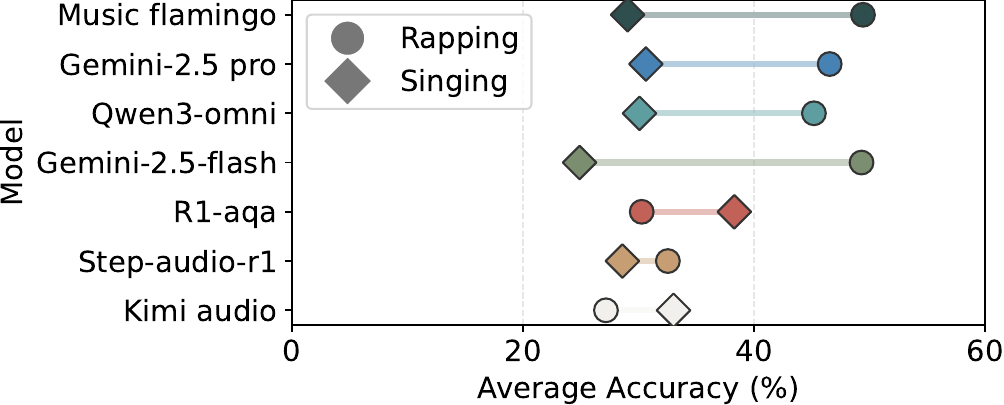}
    \caption{Model accuracy for singing versus rapping vocal styles. Most models identify rapping more accurately than singing.}
    \label{fig:collab_analysis_delivery_type}
\end{figure}

\paragraph{Can audio LMs disentangle artist roles and vocal styles over time in collaborative music?}
Collaborative music presents a complex challenge for audio LMs, with models showing lower performance compared to other task categories. 
Performance is particularly poor for fine-grained localization and quantifying the duration of individual artists’ contributions within a song.
This finding aligns with prior work on general purpose audio benchmarks that study temporal localization 
\citep{ahia2025blabbrutallylongaudio, kumar2025mmau, ma2025mmar}. 
We also observe that models are generally better at identifying rapping than singing (\autoref{fig:collab_analysis_delivery_type}). This might be due to rapping more closely resembling spoken language, which models are typically trained on.


\begin{figure*}[t] 
    \centering
    \includegraphics[width=0.95\textwidth]{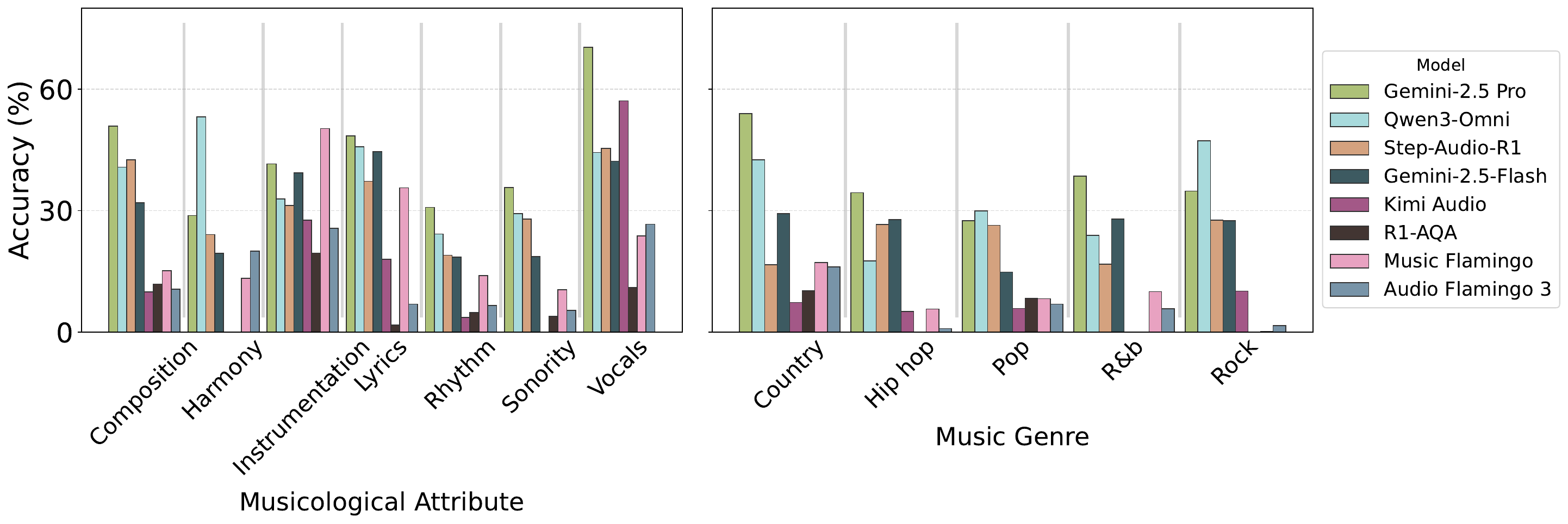}
    \caption{Model performance across musicological attributes and genres in the musicological analysis task. \textbf{(Left)} Average performance across specific attributes; while models excel in vocals and lyrics, attributes like rhythm, instrumentation, and sonority remain challenging even for frontier models. \textbf{(Right)} Performance breakdown by genre, showing that models generally perform best on Country and Rock while experiencing a noticeable dip on Hip Hop and Pop.}
    \label{fig:musicological_attribute_genre_analysis}
\end{figure*}

\paragraph{How well can audio LMs reason about musicological attributes?}
Music understanding involves reasoning over a variety of musicological attributes present in the audio, from vocals and lyrics to rhythm etc. To assess how well audio LMs capture these properties, we analyze model performance across different musicological attributes in \autoref{fig:musicological_attribute_genre_analysis}. Results vary across models and attributes. Gemini-2.5 Pro performs strongest overall, especially on Vocals, Lyrics, and Composition, while Music Flamingo excels mainly in Instrumentation but lags in Harmony. All models evidently struggle with Sonority and Rhythm, suggesting that capturing subtle timbral and temporal nuances remains a key challenge for current audio LMs, consistent with our earlier observations of temporal reasoning difficulties in audio LMs.

\paragraph{Do audio LMs benefit from source separation for structural and semantic music reasoning?}
Music recordings usually contain a mix of vocals, instruments, and background sounds, differing significantly from conventional speech. Previous work has investigated the effects of source separation---isolating individual stems; vocals, bass, drums, and other instruments from a mixed recording---on downstream music understanding tasks \citep{DeBerardinis2020MusicEmotion, Xu2014SourceSepMER, 10.1007/s10844-017-0464-5, AGRAWAL2023100254, syed2025exploitingmusicsourceseparation}, and  consistently reported performance gains when models operate only on vocal stems. We investigate whether the same benefits hold for audio LMs across the tasks in our benchmark. 

\begin{figure}[h]
    \centering
    \includegraphics[width=\columnwidth]{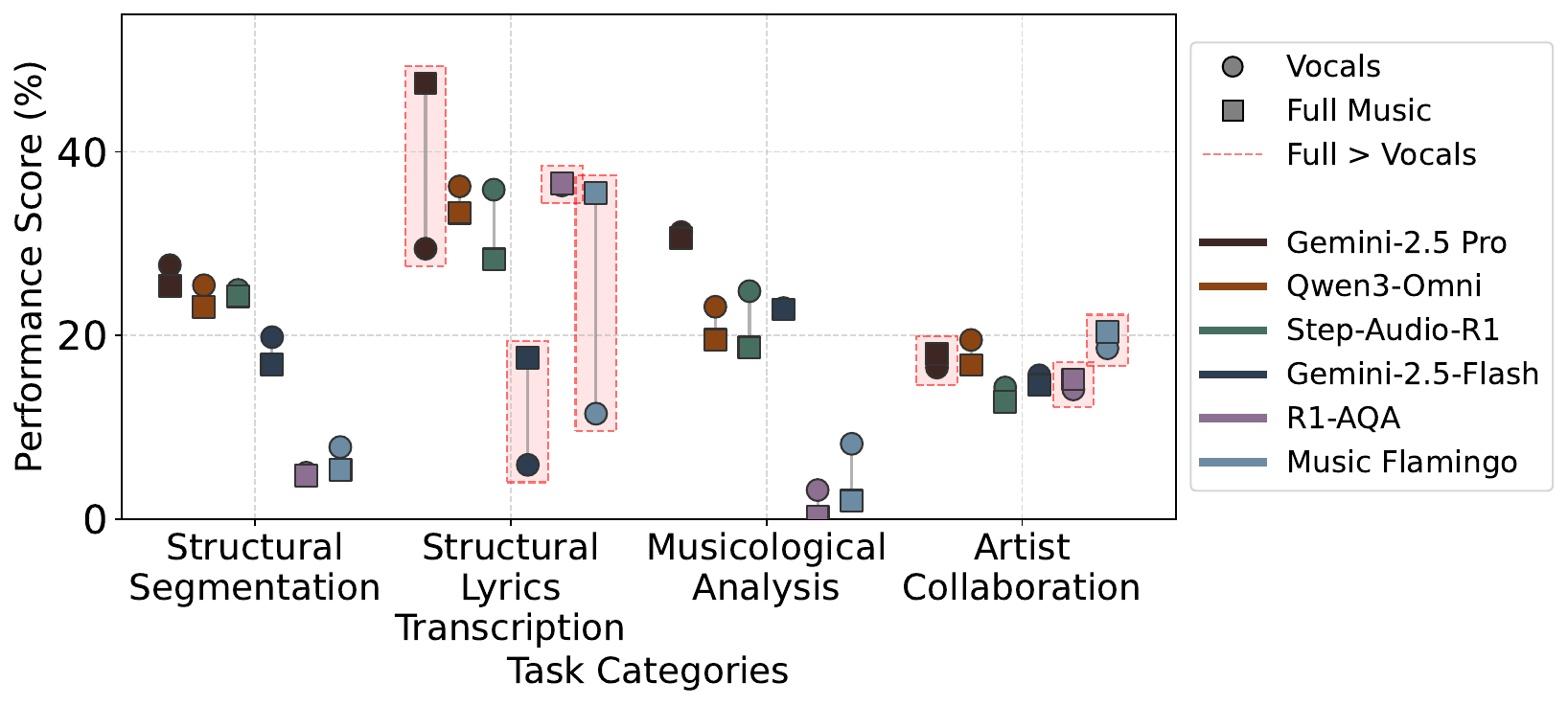}
    \caption{Source separation degrades performance on music understanding tasks. Mixed audio \textit{(squares)} consistently outperform vocal stems (circles) especially in structural lyrics transcription. }
    \label{fig:performance_full_vocals}
\end{figure}

We use the BS-RoFormer model \citep{lu2023musicsourceseparationbandsplit} to replace the mixed audio in our benchmark with vocal stems and reevaluate all models using the same configuration. As shown in \autoref{fig:performance_full_vocals}, vocal stems generally degrade performance across tasks to varying degrees. Overall, models perform worse on most tasks, indicating that mixed audio contains richer information and that source separation does not reliably improve performance for reasoning about musical structure and semantics.

\paragraph{Which musical genres challenge audio LMs the most?}
\benchmark{} contains recordings across musical genres as displayed in \autoref{fig:genre_distribution}. For this analysis, we focus on musicological task and analyze average model performance to understand the most challenging genres for audio LMs. As shown in \autoref{fig:musicological_attribute_genre_analysis}, models perform best on \textit{Country} and \textit{Rock} but relatively worse on \textit{Pop}. This suggests that widely listened to genres like Pop may be underrepresented in the training data of current audio LMs.

\paragraph{Does step‑by‑step reasoning improve music understanding performance?}
Many tasks in \benchmark{} require compositional skills and may benefit from step-by-step reasoning. While many audio LMs are not trained with hybrid reasoning modes (i.e., capable of switching between a deeper, step-by-step reasoning mode and a faster, direct response mode), we investigate how such reasoning capabilities impact performance across all tasks in our benchmark. We perform this analysis on Qwen3-Omni.\footnote{Closed-source models like Gemini Flash also have ``Thinking'' and non-``Thinking'' variants, but it is unclear if these use the same underlying model.} In our experiments, we use vLLM’s reasoning control method\footnote{\tiny \url{https://docs.vllm.ai/en/latest/features/reasoning_outputs/\#server-level-default-chat-template-kwargs}} 
to switch Qwen3-Omni between ``thinking'' and ``non-thinking modes'', performing all experiments across three independent seeds.

\noindent Our results in \autoref{tab:thinking_vs_non_thinking} indicate that enabling ``thinking'' significantly improved model performance on 6 out of 12 tasks in \benchmark{}. We observed the largest performance gains on Artist Attribution (+24.53), Gene Attribution (+17.97), and Section Lyrics Transcription (+13.33), all of which are statistically significant ($p < 0.001$) after Holm-Bonferroni correction. Tasks like Full Lyrical Transcription and (-12.77) and Gene Dominance Ranking (-2.11) show small negative performance deltas where ``thinking'' degrades performance.

\input{tables/thinking_vs_no_thinking}


\paragraph{Do Audio LMs struggle to connect memorized metadata to musical audio?}

\begin{figure*}[h]
    \centering
    \includegraphics[width=\textwidth]{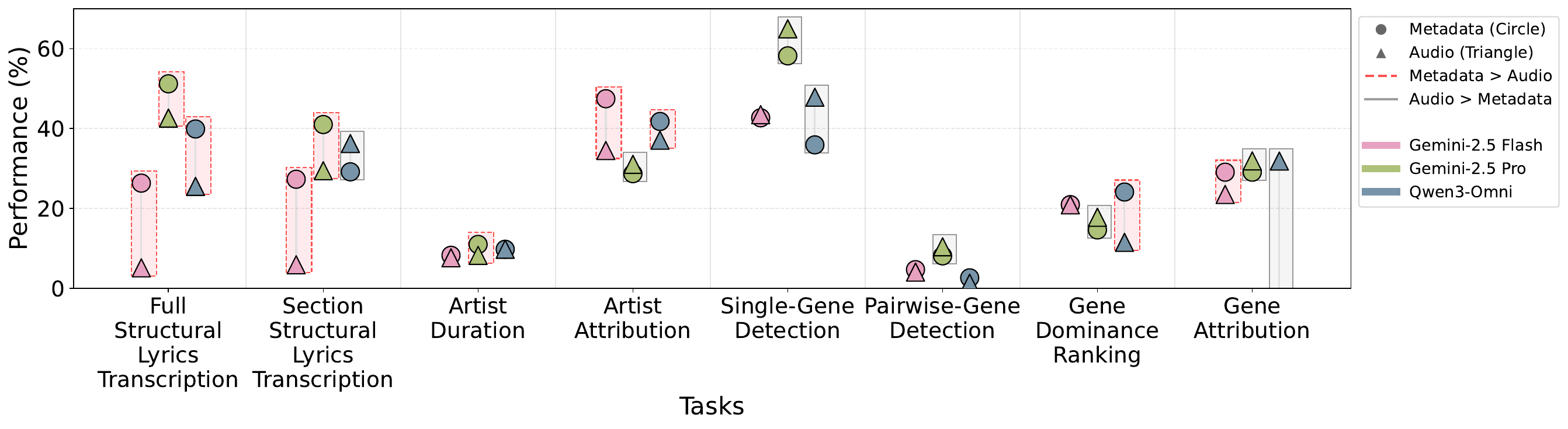}
   \caption{Performance comparison when musical audio is replaced with metadata (artist name and song title). When using only metadata, models achieve higher relative performance on lyrics transcription than on other tasks, suggesting a stronger reliance on memorized artist–song associations in the training data.}
    \label{fig:memorization_difference}
\end{figure*}

\benchmark{} contains musical recordings across genres, particularly popular songs. Given that many recent LMs are trained on large-scale web data, which may include overlapping musical content or metadata, it is important to assess the extent to which model performance reflects memorization rather than audio-based reasoning. To investigate this, we provide models only with the artist name and song title along with the question, omitting the audio. We then compare performance in this text-only condition to performance when the musical audio recording is provided. This analysis is performed on Qwen3-Omni, Gemini, and Gemini Pro, focusing specifically on the artist collaboration, musicological analysis, and structural lyrics transcription tasks. We excluded questions that require timestamp localization, since we do not provide any audio for these evaluations. 

Our results in \autoref{fig:memorization_difference} show that providing only the metadata (artist name and song title) surprisingly boosts performance on lyrics transcription tasks compared to providing the audio alone, with Gemini Flash having the largest improvement. While this improvement indicates that models can leverage memorized artist–song associations in the training data, it also highlights the challenges they face when text priors are absent, leaving substantial room for improvement. On other tasks, performance either drops or remains similar to the audio condition. This aligns with our expectations, since these tasks rely more on information present in the audio than on memorized metadata. It is still a little surprising that, even though the LM component of the audio LM might have seen the lyrics during training, the audio LMs struggles to connect this prior knowledge to the musical audio at inference.





%% file: tables/cascaded_lyrics_transcription.tex
\begin{table}[ht]
    \centering
    \scriptsize
    \newcolumntype{Y}{>{\centering\arraybackslash}X} 
    \setlength{\tabcolsep}{2pt} 
    \begin{tabularx}{\columnwidth}{lYYYY}
        \toprule
        \textbf{Setup (IWER $\uparrow)$} & \makecell[c]{\textbf{Qwen-3}\\\textbf{Omni}} & \makecell[c]{\textbf{Gemini}\\\textbf{2.5 Pro}} & \makecell[c]{\textbf{Phi-MM}\\\textbf{Instr.}} & \makecell[c]{\textbf{Gemma}\\\textbf{3n}} \\
        \midrule
        \multicolumn{5}{l}{\textbf{\textit{\color{blue} Full Lyrics Transcription (FLT)}}} \\
        Cascaded (Whisper+Audio LM) & 6.14  & 13.25 & \textbf{3.02} & 10.90 \\
        Audio LM only      & \textbf{11.38} & \textbf{42.09} & 0.90 & \textbf{14.45} \\
        \midrule
        \multicolumn{5}{l}{\textbf{\textit{\color{blue} Section Lyrics Transcription (SLT)}}} \\
        Cascaded (Whisper+Audio LM) & 23.77 & 29.69 & 0.72 & 5.51  \\
        Audio LM only      & \textbf{26.54} & \textbf{46.11} & 0.3   & \textbf{34.11} \\
        \bottomrule
    \end{tabularx}
    \caption{Using a cascaded Whisper transcript reduces audio LMs’ lyrics transcription performance compared to audio-only transcription. (Values reported in inverted WER; higher is better)}
    \label{tab:cascaded_vs_audio}
\end{table}

%% file: tables/thinking_vs_no_thinking.tex
\begin{table}[h!]
\scriptsize
\centering
\centering
\resizebox{\columnwidth}{!}{%
\begin{tabular}{lccc}
\toprule
Task & No Thinking & Thinking & $\Delta$ \\
\midrule
\multicolumn{4}{l}{\textbf{Artist Collaboration}} \\
\midrule
Artist Counting                     & 25.04 & 24.59 & -0.46 \\ 
\rowcolor{green!20} Artist Duration       & 4.59  & 8.97  & +4.38 \\ 
\rowcolor{green!20} Artist Attribution   & 11.10 & 35.63 & +24.53 \\ 
Artist Localization                 & 3.64  & 7.29  & +3.65 \\ 
\midrule
\multicolumn{4}{l}{\textbf{Musicological Analysis}} \\
\midrule
Single-Gene Detection               & 45.47 & 46.27 & +0.80 \\ 
Pairwise-Gene Detection             & 3.30  & 4.91  & +1.61 \\ 
Gene Dominance Ranking              & 17.79 & 15.68 & -2.11 \\ 
\rowcolor{green!20} Gene Attribution       & 13.00 & 30.97 & +17.97 \\ 
\midrule
\multicolumn{4}{l}{\textbf{Structural Segmentation}} \\
\midrule
\rowcolor{green!20} Full Structural Segmentation    & 24.05 & 27.19 & +3.14 \\ 
\rowcolor{green!20} Section Structural Segmentation & 15.06 & 19.84 & +4.77 \\ 
\midrule
\multicolumn{4}{l}{\textbf{Structural Lyric Transcription}} \\
\midrule
Full Structural Lyric Transcription    & 28.08 & 15.31 & -12.77 \\ 
\rowcolor{green!20} Section Structural Lyric Transcription & 27.86 & 41.19 & +13.33 \\ 
\bottomrule
\end{tabular}}
\caption{Performance comparison of Qwen3-Omni with and without "thinking". Tasks where enabling "thinking" led to statistically significant improvements (p $<$ 0.001  after Holm-Bonferroni correction across three seeds) are highlighted in green.}
\label{tab:thinking_vs_non_thinking}
\end{table}

%% file: 06_related_work.tex
\section{Related Work}
Existing benchmarks that include music-related tasks are broad audio benchmarks designed to evaluate general audio understanding in large audio LMs. Examples include BLAB \citep{ahia2025blabbrutallylongaudio},  MMAU \citep{sakshi2025mmau}, MMAU-Pro \citep{kumar2025mmau}, MMAR \citep{ma2025mmar}, and AIR-Bench \citep{yang2024airbenchbenchmarkinglargeaudiolanguage}. Music is treated as one of many audio domains in some of them, and the evaluation primarily emphasizes on general acoustic understanding rather than music-specific reasoning.
MuChoMusic \citep{weck2024muchomusic} focuses exclusively on musical understanding. It constructs multiple-choice questions using MusicCap and the Song Describing Dataset (SDD) \citep{manco2023thesong}. However, MusicCaps is limited to 10-second excerpts and SDD to two-minute clips, preventing evaluation of long-range musical structure.
CMI‑Bench \citep{ma2025cmibenchcomprehensivebenchmarkevaluating}  does include open‑ended tasks, but it is centered on classical music information retrieval (MIR) tasks and reframes traditional MIR annotations into instruction following formats rather than long‑form musical reasoning. MusicQA  \citep{liu2023musicunderstandingllamaadvancing} introduced as part of the Music Understanding LLaMA work, provides question answering and captioning capabilities for music, but its task diversity is constrained to general music knowledge and generation rather than comprehensive reasoning across extended music contexts. Our \benchmark{} is explicitly designed to evaluate both semantic and structural musical understanding and reasoning over extended temporal contexts.

%% file: 07_conclusion.tex
\section{Conclusion}
In this paper, we introduce \benchmark{}, designed to evaluate music understanding and reasoning in AI systems. \benchmark{} covers four categories—\textit{Structural Segmentation}, \textit{Lyric Transcription}, \textit{Musicological Analysis}, and \textit{Artist Collaboration} —across 12 tasks requiring both structural and semantic reasoning. We evaluate 14 open-source and frontier audio LMs, including models trained with hybrid thinking, and find that even the best-performing model achieves only 26\% average performance. Our analysis shows that state-of-the-art models struggle on higher-level reasoning tasks such as structural segmentation and artist collaboration, while performing relatively better on lyric transcription. \benchmark{} highlights model strengths and gaps to guide future development of audio LMs  with strong long-form music understanding and reasoning capabilities.

\section{Limitations}
The songs in \benchmark{} is predominantly in English or features English-speaking artists. We acknowledge that the scope of this work is centered on English-language content, as many existing audio LLMs are primarily trained on English datasets. The performance of models on this benchmark may not fully generalize to music in multilingual or multicultural settings. We leave this exploration for future work. 
Our data curation process draws on existing music resources. To ensure quality, we applied rigorous filtering so that all data comes from credible sources and is annotated by professional musicologists.


\section*{Ethical Statement}
\benchmark{} is sourced from Creative Commons–licensed music resources, with audio obtained from YouTube. Due to licensing restrictions, we do not plan to distribute the music audio files directly. Instead, when we release the dataset, we will provide links to the YouTube urls for each track, allowing users to access the audio legally while respecting copyright. While we have curated the dataset to focus on publicly available content, users should be aware that it contains works from diverse artists ; we encourage responsible and respectful use. To promote transparency and reproducibility, we will make our benchmark publicly available, along with associated evaluation metrics and the data curation framework, allowing the research community to contribute and build upon our work.

%% file: appendix.tex
\section{Dataset Details}
\subsection{Dataset Statistics}
\begin{figure}[htbp]
    \centering
    \includegraphics[width=\columnwidth]{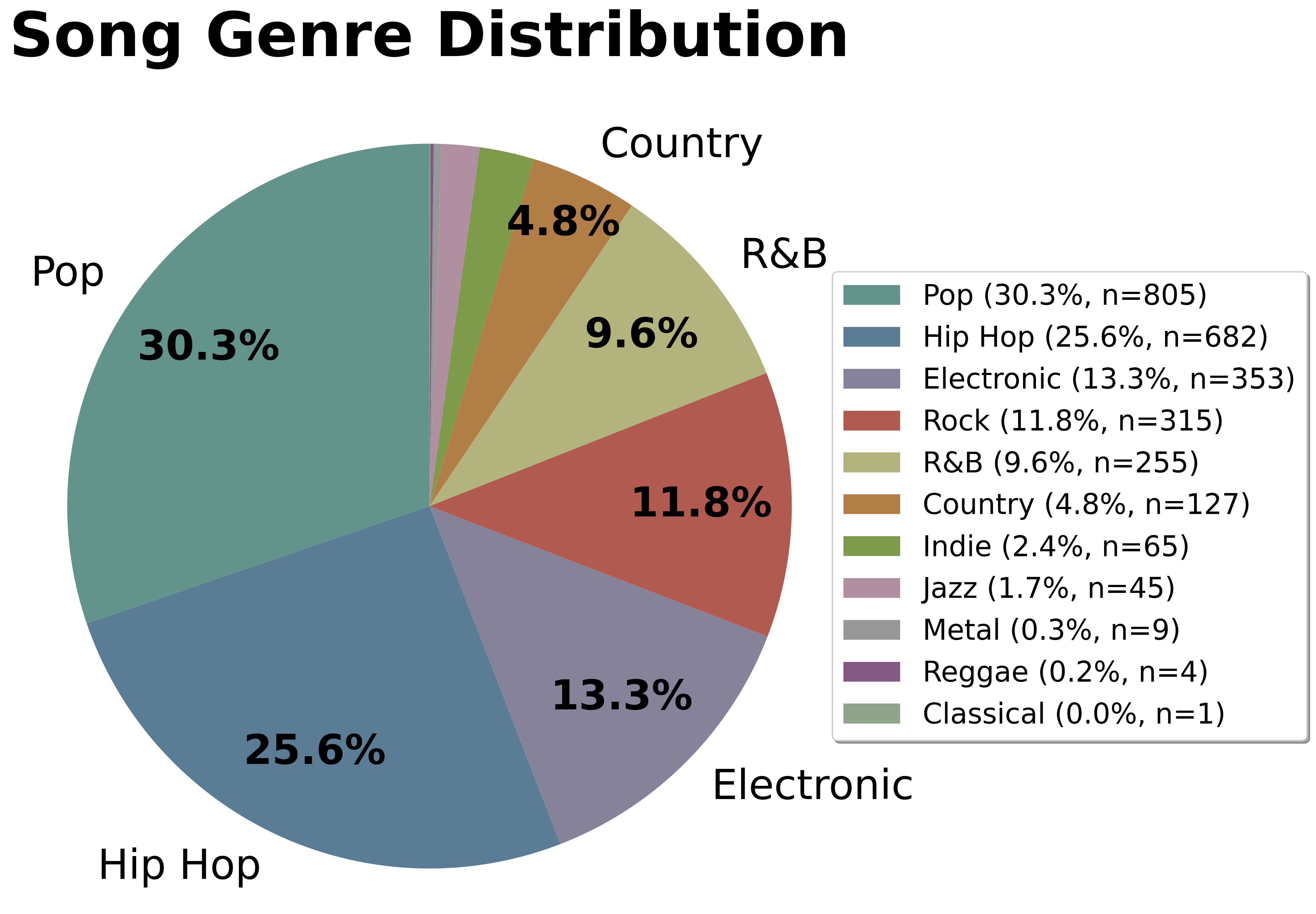}
    \caption{Distribution of musical genres across songs in \benchmark{}}
    \label{fig:genre_distribution}
\end{figure}

We provide a comprehensive breakdown of data statistics across tasks in \benchmark{} in \autoref{tab:data_stats}. In \autoref{fig:genre_distribution}, we present distribution of musical genres across the dataset, which is led by Pop (30.3\%) and Hip Hop (25.6\%), followed by a diverse range of categories including Electronic, Rock, and R\&B.

\input{tables/data_stats_new}

\subsection{Prompts for all tasks}
In \autoref{tab:example_questions}, we present the specific prompts and response formatting instructions for all tasks in the benchmark.

\input{tables/new_questions}

\newpage
\section{Task-Specific Results}
In \autoref{sec:results} in the main paper, we present an overview of our key findings and discuss their implications for audio LMs’ reasoning across musical structure, lyrics, artist roles, and genres. Here, we present finegrained results for each task.

\subsection{Structural Segmentation}
\begin{figure}[h]
    \centering
    \includegraphics[width=\columnwidth]{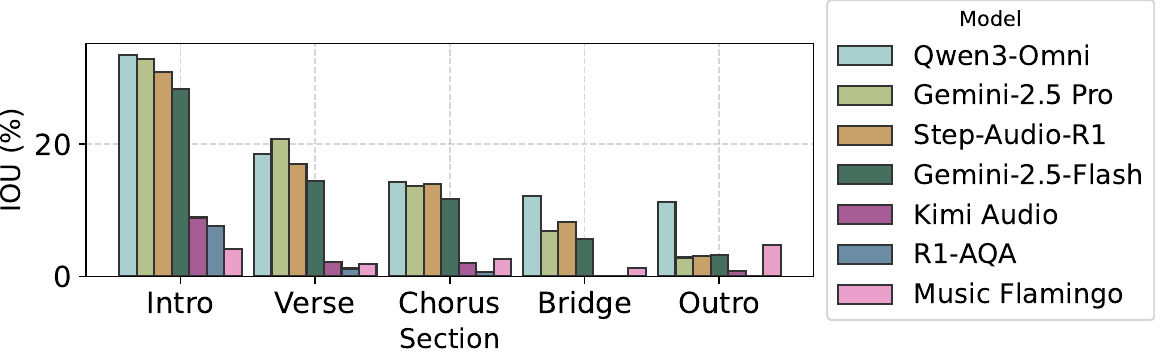}
    \caption{Section-wise performance for structural segmentation. Performance is highest for earlier sections and gradually decreases toward the end of the recording.}
    \label{fig:structural_seg_bar}
\end{figure}

For structural segmentation tasks, our results show that models generally perform better when localizing all sections within a song (Full Segmentation) compared to localizing a single, isolated section (Sectional Segmentation). Across all evaluated models, the average absolute normalized score difference between these two tasks is 7.63\%.
In \autoref{fig:structural_seg_bar}, we present the average model performance per section across the top performing models. We observe that models are more capable of segmenting sections that occur earlier in the song, such as the intro and verses (1st verse). There is a gradual decline in performance as the songs progress toward the outro section.

\begin{figure*}[htbp]
    \centering
    \includegraphics[width=\textwidth]{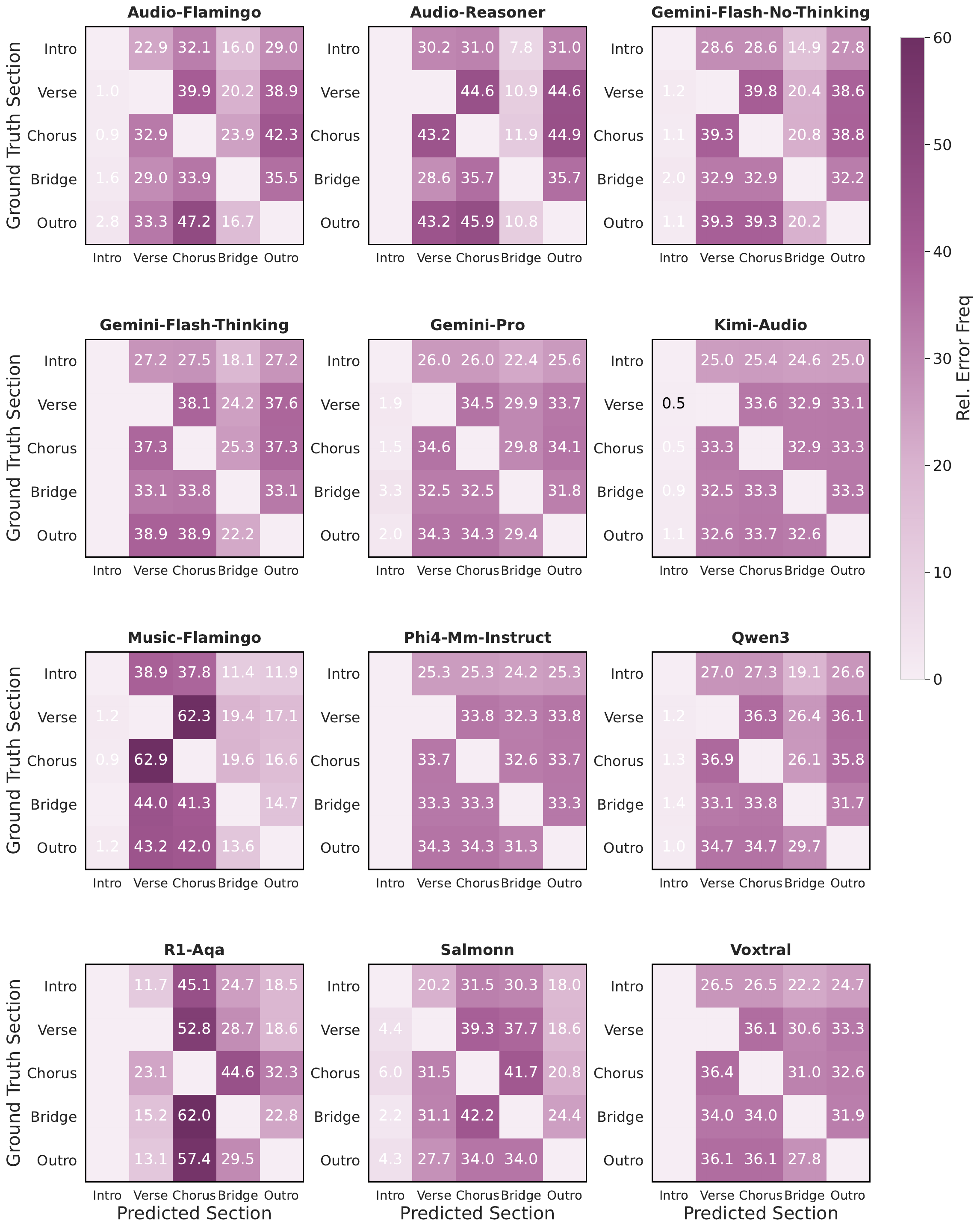}
    \caption{Section prediction errors across models
for structural segmentation, highlighting frequent
confusion between choruses and verse sections. }
    \label{fig:structural_segmentation_confusion_all_models}
\end{figure*}

\autoref{fig:structural_segmentation_confusion_all_models} shows a confusion matrix across all models. We observe that the verse and chorus are most commonly mistaken for each other.  We do not expect acoustic information alone to distinguish verses from choruses across songs; cues such as repeated melody and lyrics in choruses should be more informative. We speculate this confusion is due to audio LMs not being explicitly trained to understand musical structure across long range dependencies.

\subsection{Structural Lyrics Transcription}
For this task, we observe trends broadly consistent with structural segmentation: models perform substantially better on full than on section-level transcription. Across all evaluated models, the average normalized performance difference between full-song and section-level transcription is 4.35\% (measured via IWER \autoref{sec:metrics_description}).

While global performance patterns resemble those seen in structural segmentation, we note a distinct gap in the \textit{Intro} sections. We hypothesize that this arises because \textit{Intros} and \textit{Outros} often lack lyrics; as a result, models are more likely to produce insertion errors, possibly hallucinating words or repeating phrases from other sections.

\begin{figure*}[htbp]
    \centering
    \includegraphics[width=\textwidth]{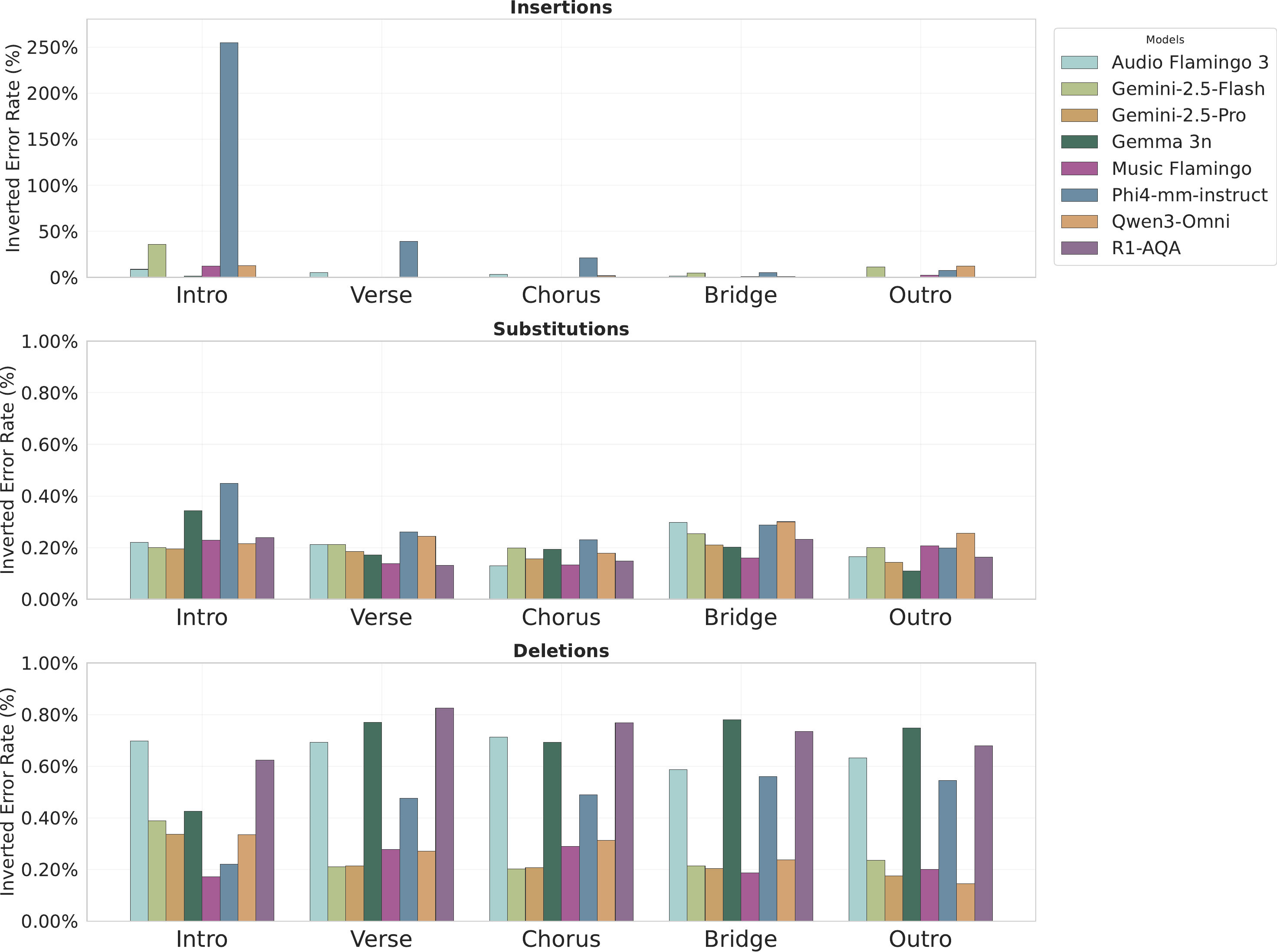}
    \caption{Analysis of inverted error rates by song section. Breakdown of Inverted \textbf{insertion}, \textbf{substitution}, and \textbf{deletion} error rates across sections. Note the significantly higher scale for \textbf{insertion} errors \textbf{(up to 250\%)} compared to \textbf{substitutions} and \textbf{deletions} (under 1.00\%), particularly in the \textit{Intro}  section.} 
    \label{fig:finegrained_error_bar}
\end{figure*}

\begin{figure*}[htbp]
    \centering
    \includegraphics[width=\textwidth]{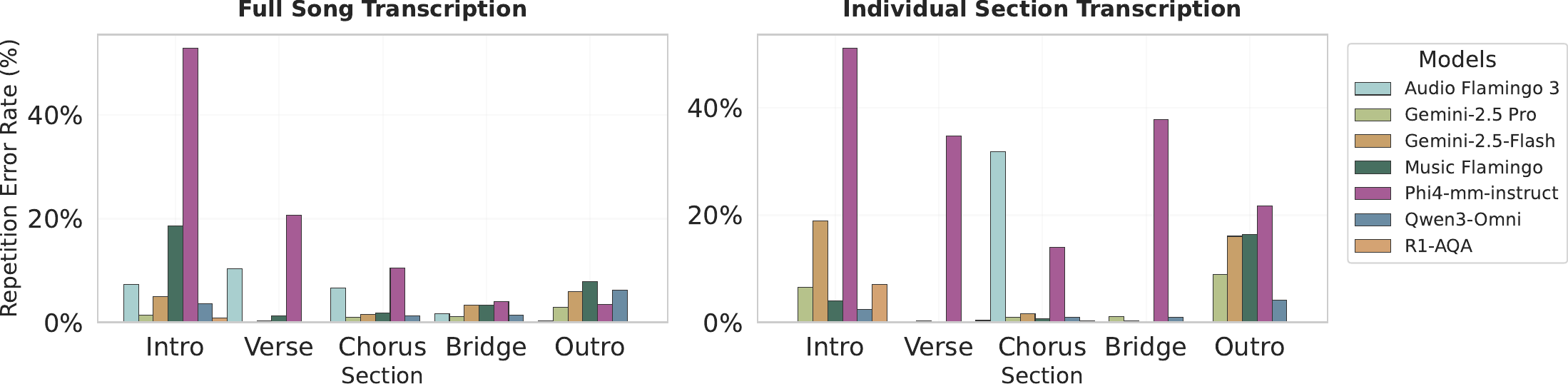}
    \caption{Repetition error rates by section for Full and Section transcription. We observe higher repetition errors (phrase hallucinations) in the \textit{Intro} across both tasks, with Section transcription consistently exhibiting higher repetition rates than Full transcription. }
    \label{fig:repetition_error_bar}
\end{figure*}

To investigate this, we analyze fine-grained WER and error types. We find that models produce more deletions in \textit{Verses} and more insertions in \textit{Intros} and \textit{Outros} (Appendix, \autoref{fig:finegrained_error_bar}), contributing to lower transcription performance in these sections. We further characterize this behavior by quantifying repeated phrases in the predicted transcription, using
\[
\frac{\sum_{\phi \in \text{repeats}} \max(0, c_\phi - r_\phi) \cdot |\phi|}{|\text{ref\_words}| + k},
\] 
where $\phi$ is a repeated phrase, $c_\phi$ is its count in the prediction, $r_\phi$ is its count in the reference, $|\phi|$ is the phrase length in words, $|\text{ref\_words}|$ is the total number of words in the reference, and $k$ is a smoothing constant to prevent inflated scores for very short references. As shown in \autoref{fig:repetition_error_bar}, repetition errors are indeed most prevalent in \textit{Intro} and \textit{Outro} sections on both full and section-level transcriptions, supporting our hypothesis.

\subsection{Musicological Analysis}
For this task, we observe the highest performance on Single-Gene Detection where models are prompted to select a single musicological attribute that is strongly dominant in the song  (e.g., presence of a rhythmic groove) out of four options. Performance drops substantially in the Pair-Gene Detection task, where models must select a pair of attributes from eight options, making the task more challenging; even  Gemini achieves only around 10\% accuracy. Performance is similarly low for Gene Dominance Ranking, which requires ordering attributes by dominance. However, results improve for Gene Attribution, where models operate at a higher level by selecting a single song with a dominant attribute from a set of four songs.


\subsection{Artist Collaboration}
\begin{figure}[h]
    \centering
    \includegraphics[width=\columnwidth]{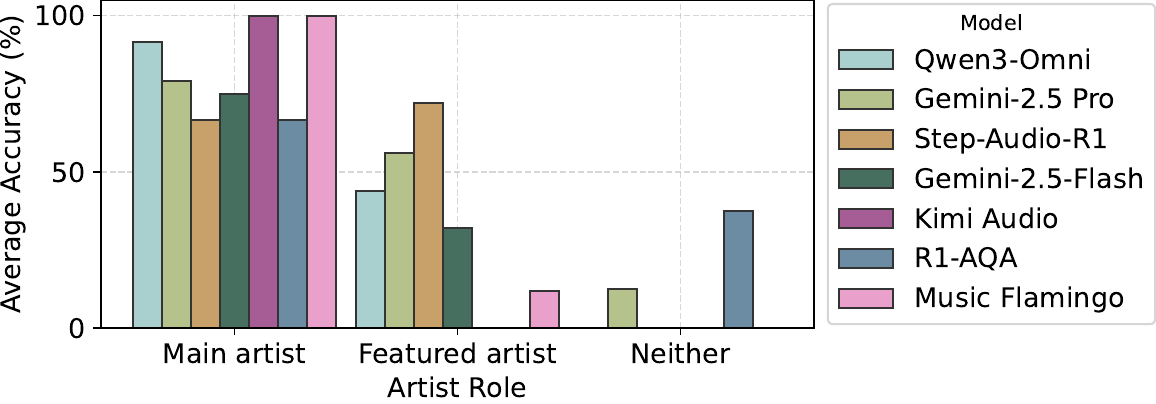}
    \caption{Model accuracy across different artist roles.
Accuracy is highest for main artists, lower for featured
artists, and lowest for artists in neither category.}
    \label{fig:collab_analysis_role_accuracy}
\end{figure}

Across all task categories, this category appears to be the most challenging. While overall performance is low, models consistently perform better on artist counting and artist attribution than others. 
These tasks share an underlying focus on recognizing vocal style, artist roles, and temporal localization. Previous benchmarks have reported audio LMs lacking temporal localization skills \citep{ahia2025blabbrutallylongaudio, kumar2025mmau, ma2025mmar}, so this result is not surprising. 
In \autoref{fig:collab_analysis_role_accuracy}, we analyze tasks that require identifying artist roles within a song at varying levels of granularity. Models are generally more accurate at identifying main artists than featured artists, and perform worst on cases where artists are neither main nor featured. 

For vocal style delivery, we analyze performance differences between identifying singing and rapping (\autoref{fig:collab_analysis_delivery_type}). Most models achieve higher accuracy on questions related to rapping, with the widest gap for Gemini Flash. 

Finally, we quantified prediction errors across artist counting and duration-related tasks (see Appendix \autoref{fig:rtist_count_error} and \ref{fig:artist_duration_error}). Gemini Pro achieves near-zero error, while Audio-Flamingo, Audio-Reasoner, and Phi-mm Instruct tend to underestimate durations, and Gemini Flash and R1 AQA overestimate. Similar trends appear in artist counting, with most models underestimates artist counts on the full songs. Extreme outliers (e.g., Gemma, Qwen) are omitted from the plots due to hallucinated errors exceeding $10^7$ seconds that would dominate the scale of the plot.

\begin{figure}[htbp]
    \centering
\includegraphics[width=0.6\columnwidth]{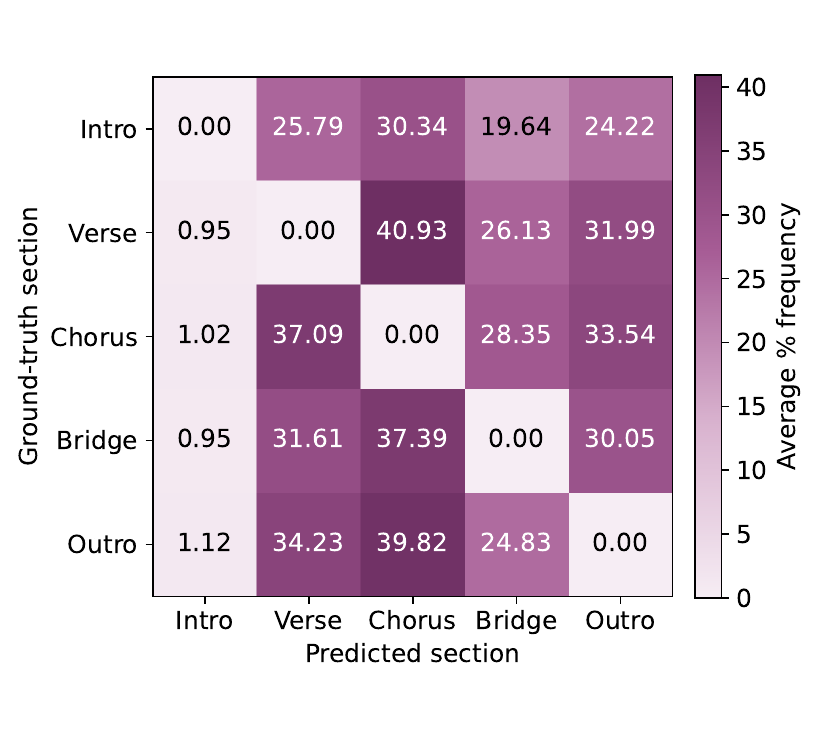}
    \caption{Average section prediction errors across models for structural segmentation, highlighting frequent confusion between choruses and verse sections.}
    \label{fig:structural_seg_heatmap}
\end{figure}

\input{tables/full_music_raw_scores}

\begin{figure*}[htbp]
    \centering
    \includegraphics[width=\textwidth]{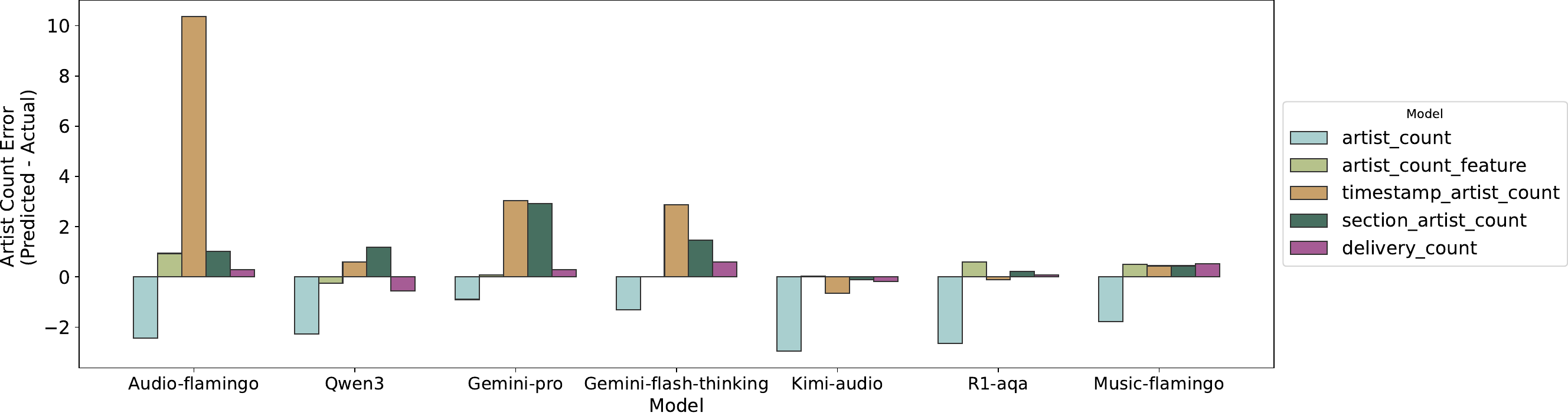}
    \caption{Model performance across artist counting subtasks. The y -- axis shows error (Predicted -- Actual), indicating overestimation (positive) or underestimation (negative) of artist and delivery counts.}
    \label{fig:rtist_count_error}
\end{figure*}

\begin{figure*}[htbp]
    \centering
    \includegraphics[width=\textwidth]{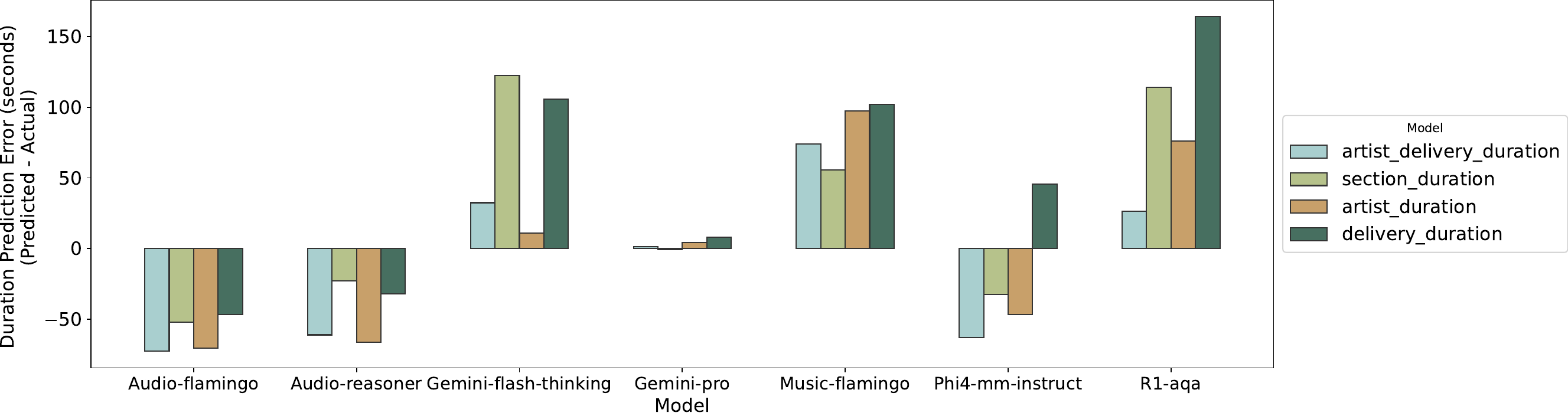}
    \caption{Temporal prediction errors (seconds) across artist duration tasks. The y -- axis shows error (Predicted -- Actual), indicating overestimation (positive) or underestimation (negative) of duration. }
    \label{fig:artist_duration_error}
\end{figure*}

%% file: tables/data_stats_new.tex
\begin{table}[h]
\centering
\small
\setlength{\tabcolsep}{3pt} 
\begin{tabular}{llccccc}
\toprule
\textbf{Cat.} & \textbf{Task} & \textbf{\# Q} & \textbf{\# S} & \textbf{Avg.} & \textbf{Min.} & \textbf{Max.} \\
\midrule
\multirow{3}{*}{Struct.} 
 & FSS & 74 & 74 & 3.06 & 1.70 & 10.78 \\
 & SSS & 224 & 224 & 3.47 & 1.93 & 9.31 \\
 & Overall & 298 & 298 & 3.37 & 1.70 & 10.78 \\
\midrule
\multirow{3}{*}{Lyrics} 
 & FSLT & 665 & 665 & 3.73 & 1.60 & 11.72 \\
 & SSLT & 170 & 170 & 3.97 & 2.17 & 8.44 \\
 & Overall & 884 & 884 & 3.78 & 1.60 & 11.72 \\
\midrule
\multirow{5}{*}{Music.} 
 & SGD & 335 & 335 & 3.15 & 1.86 & 9.23 \\
 & PGD & 147 & 147 & 3.77 & 0.82 & 6.37 \\
 & GDR & 33 & 33 & 3.60 & 2.58 & 4.95 \\
 & GA & 94 & 94 & 14.14 & 9.77 & 19.44 \\
 & Overall & 609 & 484 & 5.57 & 0.82 & 19.44 \\
\midrule
\multirow{5}{*}{Collab.} 
 & Cnt & 229 & 229 & 3.92 & 2.37 & 6.97 \\
 & Dur & 327 & 327 & 3.83 & 1.34 & 6.97 \\
 & Loc & 247 & 247 & 3.83 & 2.37 & 6.13 \\
 & Attr & 64 & 64 & 3.88 & 1.34 & 6.97 \\
 & Overall & 867 & 327 & 3.83 & 1.34 & 6.97 \\
\bottomrule
\end{tabular}
\caption{Data statistics of each task.}
\label{tab:data_stats}
\end{table}

%% file: tables/new_questions.tex
\begin{table*}[t]
\centering
\scriptsize
\renewcommand{\arraystretch}{1.2}
\begin{tabularx}{\textwidth}{l p{2.5cm} p{2.5cm} X}
\toprule
\textbf{Category} & \textbf{Task} & \textbf{Subtasks} & \textbf{Example Question \& Expected Output Format} \\
\midrule
\multirow{2}{*}{\shortstack[l]{Structural \\ Segmentation}} 
& Full Structural Segmentation & -- & Divide the track into sections corresponding to any of these categories: Verse, Chorus, Bridge, Intro, Outro, Pre-Chorus, Post-Chorus. For each section, provide the following: section title, start and end timestamps of the entire section. \par \textbf{Format:} \texttt{[ \{"section": "<title>", "start": <ts>, "end": <ts> \}, ... ]} \\ \cmidrule{2-4}
& Section Structural Segmentation & -- & Given the song, provide the start and end timestamp (in seconds with milliseconds) of all Verse sections. \par \textbf{Format:} \texttt{[ \{"start": <ts>, "end": <ts> \}, ... ]} \\ 
\midrule

\multirow{2}{*}{\shortstack[l]{Lyrics \\ Transcription}} 
& Full Structural Lyrics Transcription & -- & You are provided with the structural segmentation of a song. Please transcribe the lyrics for each section. \par \textbf{Format:} \texttt{[ \{"section": "<section1>", "lyrics": "..." \}, ... ]} \\ \cmidrule{2-4}
& Section Structural Lyrics Transcription & -- & Could you provide the lyrics for the all verse sections section only? The transcription of each occurrence of the given section should be treated as a separate item in a list. \par \textbf{Format:} \texttt{[ \{"section": "section\_name", "lyrics": "..." \} ]} \\
\midrule

\multirow{4}{*}{\shortstack[l]{Musicological \\ Analysis}} 
& Single-Gene Detection & -- & Given the following musical properties, which one is most dominant in the song? Answer Choices: Compound Meter, Percussion, Vocal Grittiness, Synthesizer. \par \textbf{Format:} Respond with only the name of the musicological feature. \\ \cmidrule{2-4}
& Pairwise-Gene Detection & -- & Which pair of features is most strongly present in the track? \par \textbf{Format:} Respond with only a list of two names. Example: \texttt{['<property1>', '<property2>']} \\ \cmidrule{2-4}
& Gene Attribution & -- & Which song features the highest Vocal Register? (Property measures pitch range of lead singer). \par \textbf{Format:} Respond with only the number (1, 2, 3, or 4) corresponding to the song. \\ \cmidrule{2-4}
& Gene Dominance Ranking & -- & Rank the following rhythmic patterns and feels from least to most prominent. \par \textbf{Format:} Respond with only a list of the names ranked. Example: \texttt{['<prop1>', '<prop2>']} \\
\midrule

\multirow{10}{*}{\shortstack[l]{Artist \\ Collaboration}} 
& \multirow{4}{*}{Artist Counting} & Standard / Featured & What is the number of artists performing? / How many performers are featured artists? \\ \cmidrule{3-4}
& & Vocal Delivery & In this song, what number of artists are performing rap sections? \\ \cmidrule{3-4}
& & Section / Temporal & How many artists are performing in the Post-chorus? / From 211 to 249 seconds? \par \textbf{Format (All Counting):} Respond with only the number. \\ \cmidrule{2-4}
& \multirow{3}{*}{Artist Duration} & Target / Vocal & What is the total duration of performance by artist 2? / How many seconds are devoted to rapping? \\ \cmidrule{3-4}
& & Artist Delivery / Section & Total duration of rrapped parts by artist 1? / How long do Bridge sections last? \par \textbf{Format (All Duration):} Respond with only the number in seconds with milliseconds. \\ \cmidrule{2-4}
& Artist Localization & -- & At what second does artist 2 first start performing in this song? \par \textbf{Format:} Respond with only the timestamp number in seconds with milliseconds. \\ \cmidrule{2-4}
& \multirow{3}{*}{Artist Attribution} & Vocal Comparison & Do the featured artist in the first Verse and the main artist in the second Verse use the same type of vocal delivery? \par \textbf{Format:} Respond with only: \texttt{same} or \texttt{different}. \\ \cmidrule{3-4}
& & Section Style / Role & For the first Chorus, is the artist rapping, singing, or neither? Is it the main or featured artist? \par \textbf{Format:} \texttt{Rapping}, \texttt{Singing}, \texttt{Neither} OR \texttt{Main artist}, \texttt{Featured Artist}, \texttt{Neither}. \\
\bottomrule
\end{tabularx}
\caption{Tasks, example questions, and specific formatting instructions for all categories.}
\label{tab:example_questions}
\end{table*}

%% file: tables/full_music_raw_scores.tex
\begin{table*}[t]
\rmfamily
\centering
\setlength{\tabcolsep}{3pt}
\adjustbox{max width=\textwidth}{
\begin{tabular}{l l
S[table-format=2.2] S[table-format=2.2] S[table-format=2.2] S[table-format=2.2]
S[table-format=2.2] S[table-format=2.2]
S[table-format=2.2] S[table-format=2.2] S[table-format=2.2] S[table-format=2.2]
S[table-format=2.2] S[table-format=2.2]
}

& & \multicolumn{4}{c}{\textbf{\BroadCategory{Musicological Analysis}}} 
 & \multicolumn{2}{c}{\textbf{\BroadCategory{Structural Segmentation}}}
 & \multicolumn{4}{c}{\textbf{\BroadCategory{Artist Collaboration}}}
 & \multicolumn{2}{c}{\textbf{\BroadCategory{Lyrics Transcription}}}   \\
\cmidrule(lr){3-6} \cmidrule(lr){7-8} \cmidrule(lr){9-12} \cmidrule(lr){13-14}

\textbf{Model} & \textbf{Size}
& {SGD} & {PGD} & {GDR} & {GA} 
& {Full} & {Section} 
& {Cnt.} & {Dur.} & {Loc.} & {Attr.} 
& {Full} & {Section} \\

\midrule
Gemini-2.5 Pro & - & 73.73 & 10.88 & 21.20 & 48.90 & 32.34 & 22.98 & 25.77 & 8.25 & 3.12 & 56.66 & 1.35 & 2.40 \\
Qwen3-Omni & 30B & 60.90 & 6.16 & 15.20 & 48.90 & 28.88 & 22.07 & 18.54 & 9.78 & 9.38 & 63.66 & 2.92 & 1.76 \\
Step-Audio-R1 & - & 57.91 & 10.27 & 18.18 & 48.94 & 32.19 & 17.73 & 19.34 & 6.42 & 7.81 & 53.13 & 3.57 & 3.61 \\
Gemini-2.5-Flash & - & 57.61 & 6.82 & 24.20 & 42.60 & 24.47 & 15.15 & 20.40 & 7.66 & 3.12 & 56.81 & 18.51 & 15.97 \\
Music Flamingo & 8B & 49.25 & 4.76 & 0.00 & 25.50 & 13.07 & 2.59 & 33.54 & 4.27 & 4.69 & 56.97 & 5.36 & 7.72 \\
R1-AQA & 8.2B & 34.03 & 0.00 & 0.00 & 30.90 & 5.91 & 4.16 & 35.03 & 0.92 & 0.00 & 50.84 & 1.50 & 1.75 \\
Gemma 3 & 5B & 35.22 & 4.76 & 3.03 & 31.91 & 8.92 & 4.35 & 12.25 & 0.30 & 1.56 & 40.80 & 1.37 & 1.70 \\
Audio Flamingo 3 & - & 44.20 & 8.18 & 0.00 & 30.90 & 7.03 & 0.01 & 21.92 & 1.84 & 0.00 & 49.80 & 4.68 & 2.76 \\
Voxtral-Small-24B & 24B & 37.61 & 8.84 & 9.12 & 17.00 & 32.16 & 8.84 & 9.09 & 2.76 & 0.00 & 38.52 & 2.78 & 15.46 \\
SALMONN-7B & 7B & 25.37 & 1.36 & 0.30 & 28.70 & 3.18 & 0.12 & 27.32 & 0.61 & 0.00 & 34.86 & 7.61 & 20.59 \\
Kimi Audio & 7B & 37.61 & 4.76 & 0.00 & 0.00 & 13.06 & 2.64 & 19.12 & 0.92 & 0.00 & 52.07 & 18.33 & 54.56 \\
Phi4-mm-instruct & 5.5B & 33.73 & 1.36 & 9.12 & 21.30 & 2.71 & 0.16 & 23.59 & 3.67 & 0.00 & 38.51 & 79.29 & 364.54 \\
Mellow & 167M & 0.30 & 0.00 & 0.00 & 0.00 & 0.00 & 0.00 & 0.00 & 2.14 & 0.00 & 0.00 & 19.02 & 3.46 \\
Audio-Reasoner & - & 33.73 & 3.40 & 6.06 & 30.85 & 6.63 & 2.91 & 18.47 & 3.06 & 0.00 & 2.31 & {-} & {-} \\
\bottomrule
\end{tabular}
}
\caption{Raw Performance scores \textbf{(before normalization)} of evaluated models on \benchmark{} across musicological analysis (Single-Gene Detection—SGD, Pairwise-Gene Detection—PGD, Gene Dominance Ranking—GDR, and Song-Level Gene Detection—Song-GD), structural segmentation, artist collaboration, and lyrics transcription tasks. The metric for Musicological Analysis and Artist Collaboration task is Exact Match accuracy (\%). The metric for Structural Segmentation is IOU (\%). The metric for Lyrics Transcription is WER (not in percentages).}
\label{tab:raw_scores_full_music}
\end{table*}